\documentclass{aa} 
\usepackage{txfonts}
\usepackage{graphicx}
\newcommand{\sech}{\textrm{sech}} 
\newcommand{\E}{\cdot 10^} 
\newcommand{\ud}{\mathrm{d}} 
\newcommand{\pG}{p^{}_{\rm\tiny G}} 
\newcommand{\pV}{p^{}_{\rm\tiny V}} 
\newcommand{\cD}{c^{}_\Delta} 
\newcommand{\kB}{k^{}_{\rm\tiny B}} 
\newcommand{\DCD}{{\rm\tiny DCD}} 
\newcommand{\TLS}{{\rm\tiny TLS}^{}} 
\newcommand{\IR}{{\rm\tiny IR}^{}} 
\newcommand{\mic}{\,{\rm \mu m} } 
\newcommand{\Tdust}{T_{\rm d}}        
\newcommand{\lcor}{l_{\rm c}} 
\newcommand{\aphon}{a_{\rm phon}} 
\newcommand{\kappaphon}{\kappa_{\rm phon}} 
\newcommand{\alphaphon}{\alpha_{\rm phon}} 
\newcommand{\Fphon}{F_{\rm phon}} 
\newcommand{\alphaDCD}{\alpha_{\rm DCD}} 
\newcommand{\betaphon}{\beta_{\rm phon}} 

\begin{document}
\title{Far-infrared to millimeter astrophysical dust emission}
\subtitle{I: A model based on physical properties of amorphous solids}
\author{C. Meny \inst{1} \and
    V. Gromov \inst{1,} \inst{2} \and
    N. Boudet \inst{1}\and
    J.-Ph. Bernard \inst{1}\and
    D. Paradis \inst{1}\and
    C. Nayral \inst{1}}
\offprints{Claude.Meny@cesr.fr}
\institute{Centre d'Etude Spatiale des Rayonnements, CNRS,
              9, Avenue du Colonel Roche, 31028 Toulouse, France\\
              email : meny@cesr.fr , vgromov@iki.rssi.ru
\and
              Space Research Institute, RAS, 
              84/32 Profsoyuznaya, 117810 Moscow, Russia\\}
\date{Received Juin 2006 / Accepted november 2006 ?}

\maketitle
%

\abstract
{}
{We propose a new description of astronomical dust emission in the
spectral region from the Far-Infrared to millimeter wavelengths.}
{Unlike previous classical models, this description explicitly
incorporates the effect of the disordered internal structure of
amorphous dust grains. The model is based on results from solid
state physics, used to interpret laboratory data.  The model takes
into account the effect of absorption by Disordered Charge
Distribution, as well as the effect of absorption by localized Two
Level Systems.}
{We review constraints on the various free parameters of the model
from theory and laboratory experimental data.  We show that, for
realistic values of the free parameters, the shape of the emission
spectrum will exhibit very broad structures which shape will change
with the temperature of dust grains in a non trivial way.  The
spectral shape also depends upon the parameters describing the
internal structure of the grains.  This opens new perspectives as to
identifying the nature of astronomical dust from the observed shape
of the FIR/mm emission spectrum.  A companion paper will provide an
explicit comparison of the model with astronomical data.}
{}

\section{Introduction}\label{s:intro}

It is now well established that the Spectral Energy Distribution
(SED) from the Inter-Stellar Medium (ISM) emission from our Galaxy
and most external galaxies is dominated by thermal emission from
dust grains which spans over almost 3 orders of magnitude in
wavelengths, from the Near Infra-Red (NIR) to the millimeter
wavelength range. Several dust models have been proposed to explain
the observations. Recent dust models share many common
characteristics (e.g. Mathis at al. \cite{Mathis77}, Draine \& Lee
\cite{Draine84}, Draine \& Anderson \cite{Draine85}, Weiland et al.
\cite{Weiland86}, D\'esert et al. \cite{Desert90}, Li \& Greenberg
\cite{Li97}, Dwek et al. \cite{Dwek97}, Draine \& Li
\cite{Draine01}).  Most of them require a wide range of the dust
grain sizes spanning almost 3 orders of magnitude, from
sub-nanometer dimensions to fractions of a micron. They also require
a minimum of three components with different chemical composition in
order to explain the ultraviolet and optical extinction, along with
infrared emission.

The smallest component is needed to explain the "aromatic" emission
features at 3.3, 6.2, 7.7, 8.6 and $11.3\mic$.  It is now routinely
assigned to large aromatic molecules or Polycyclic Aromatic
Hydrocarbons (PAH) (Leger \& Puget \cite{Leger84}, Allamandola et
al. \cite{Allamandola85}, D\'esert et al. \cite{Desert90}).  PAH are
transiently heated after absorbing single UV and far-UV photons, and
cool down emitting NIR photons in vibrational transitions
representative of their aromatic structure.

A second component, composed of Very Small Grains (VSG) is required to
explain the continuum emission in the Mid Infra-Red (MIR).  Emission in
this range still requires significant temperature fluctuations of the
grains, which necessitates, under conditions prevailing in the
Inter-Stellar Medium (ISM), grains sizes in the nanometer range.  VSGs
could be composed of carbonaceous material whose absorption could also
explain the $2200\AA$ UV bump in the extinction curve (see for instance
D\'esert et al.\,\cite{Desert90}).

A third component, composed of Big Grains (BG), with sizes from
10-20\,nm to about 100\,nm, is necessary to account for the long
wavelength emission (in particular as observed in the IRAS $100\mic$ band and
above). The BG component dominates the total dust mass and the
absorption in the visible and the NIR. Observations of the
"silicate" absorption feature near $10\mic$ indicate that their
mass is dominated by amorphous silicates 
(see Kemper et al. \cite{Kemper04}). Such amorphous structure for
these dust grains is expected from the interaction with cosmic
rays which should alter the nature of the solid (Brucato et al.
\cite{Brucato04}, J\"ager et al. \cite{Jager03}), but also from
the formation processes of these BGs by aggregation of sub-sized
particles.


Recent studies attempting to explain the submillimeter observations,
revealed the need for either the
existence of an additional most massive component of very cold dust
(e.g. Reach et al. \cite{Reach95}, Finkbeiner et al.
\cite{Finkbeiner99}, Galliano et al.  \cite{Galliano03}) or for
substantial modifications of dust properties in the submillimeter
with respect to predictions of standard dust models (e.g.
Ristorcelli et al. \cite{Ristorcelli98}, Bernard et al.
\cite{Bernard99}, Stepnik et al. \cite{Stepnik03A}, Dupac et al.
\cite{Dupac01G}, Dupac et al. \cite{Dupac02}, Dupac et al.
\cite{Dupac03}).

Similarly, laboratory spectroscopic measurements of interstellar
grains analogs reveal that noticeable variations in their optical
properties in the Far Infra-red (FIR) and Millimeter (mm) wavelength
range can occur (Agladze et al. \cite{Agladze94}, Agladze et al.
\cite{Agladze96}, Mennella et al. \cite{Mennella98}, Boudet et al.
\cite{Boudet05}).
They pointed out the possible role of two phonons difference
processes, disorder-induced one-phonon processes, resonant or
relaxation processes in the presence of Two-Level Systems(TLS) in
amorphous dust (see Sect.\,\ref{Sec:TLS}).  The influence of such
Two-Level Systems on interstellar grain absorption and emission
properties was first proposed to the astronomers's community by
Agladze et al. \cite{Agladze94}. More recently, a preliminary
investigations by Boudet et al. \cite{Boudet02} showed that optical
resonant and relaxation transitions in a distribution of TLS could
qualitatively explain the anticorrelation between the temperature
and the spectral index of dust as observed by the PRONAOS balloon
experiment (e.g. Dupac et al. \cite{Dupac03}).

A precise modeling of the long wavelength dust emission is
important in order to accurately subtract foreground emission in
cosmological background anisotropy measurements, especially for
future missions for space CMB measurements (Lamarre et al.
\cite{Lamarre03}) and for surveys of foreground compact sources
(see Gromov et al. \cite{Gromov02}).
A number of comprehensive foreground analysis papers 
(Bouchet \& Gispert \cite{Bouchet99}, Tegmark et al.
\cite{Tegmark00}, Bennett et al. \cite{Bennett03}, Barreiro et al.
\cite{Barreiro04}, Naselsky et al. \cite{Naselsky05} and other) did
not pay attention to dust emissivity model accuracy. In addition,
precise modeling of the FIR/mm dust emission is also
important to derive reliable estimates of the dust mass, to trace
the structure and density of pre-stellar cold cores in molecular
clouds. Indeed the dust governs the cloud opacity, and influences
the size of the cloud fragments. In addition, the efficiency of star
formation should be related to the dust properties, which are
expected to vary between the diffuse medium and the denser cold
cores. It is thus of
importance to know how some physical
or chemical properties of dust can influence their FIR/mm emission.

For models of the submillimeter electromagnetic emission of dust,
considering the intrinsic mechanical vibrations of the grain
structure is natural since they fall in the right frequency range
(typically in the frequency range $\nu\sim10^{11}-10^{13}$\,Hz,
i.e. $\lambda\rm\sim30\mic-3\,mm$).
Observations in this spectral region potentially provides new tools for
investigating the internal structure of dust grain material.  The
internal mechanics of grains is also important for chemistry in the
ISM. The dust vibrations supply an energy sink for newly formed
molecules which are usually formed in unstable excited states.
Understanding the grain structure and intrinsic movements is necessary
for considerations regarding grain collisions, destruction and
agglomeration, including preplanetary bodies formation in circumstellar
disks.

In Sect.\,\ref{Sec:Basic}, we first recall some basics regarding the
FIR/mm dust emission and the semi-classical model of light
interaction with matter.  In Sect.\,\ref{Sec:Evidences}, we gather
evidences for spectral variations in astronomical and laboratory
measurements.  Then, we present in Sect.\,\ref{Sec:ModelPhysics} a
model, based on the intrinsic properties of amorphous materials,
taking into account absorption by Disordered Charge Distribution
as well as the effect of absorption by a distribution of TLS. In
Sect.\,\ref{s:astro-model}, we
propose a new model of the submillimeter and millimeter grain
absorption and discuss the range of plausible values for the free
parameters, based on existing laboratory data.  In
Sect.\,\ref{Sec:Discussion}, we discuss the implications of the model,
in particular regarding the expected variations of the FIR/mm emission spectra with the dust temperature.
Sect.\,\ref{Sec:Conclusion} is devoted to conclusions.  The
determination of the model parameters applicable to astronomical data
will addressed in a companion paper.

\section{Basic knowledge on the FIR/mm dust emission} 
\label{Sec:Basic}

\subsection{Interstellar dust emission and extinction}

The intensity of thermal emission from interstellar dust at temperature $\Tdust$ is
\begin{equation}
    \label{eq:I}
    I_{\omega}(\omega, T)=\epsilon_{\rm e}(\omega)\cdot B_{\omega}(\omega,\Tdust)
\end{equation}
where $I_{\omega}$ is energy flux density per unit area, angular
frequency and solid angle (or specific intensity), $\epsilon_{\rm
e}$ the dust emissivity and $B_{\omega}$ the Planck function at
angular frequency $\omega$.

According to the Kirchhoff law, the emissivity $\epsilon_{\rm e}$ is equal to
the absorptivity
\begin{equation}
  \label{eq:Kirchhoff}
  \epsilon_{\rm e}(\omega)=1-e^{-\tau(\omega)}.
\end{equation}
where the optical depth $\tau$ is related to the dust mass column
density on the line of sight $M_d$ and the dust mass opacity $\kappa$ as
\begin{equation}
  \label{eq:tau}
  \tau(\omega)=\kappa(\omega)\cdot M_d
\end{equation}

For
spherical grains of radius $a$ and density $\rho$, the dust
opacity (effective area per mass) is given by

\begin{equation}
    \label{eq:Km}
    \kappa(\omega) = \frac{3}{4\rho} \frac{Q(\omega)}{a},
\end{equation}
where the absorption efficiency $Q(\omega)=\sigma(\omega)/(\pi a^2)$
is the ratio of the absorption cross section $\sigma(\omega)$ to the
geometrical cross section of the grain $\pi a^2$. The grain
equilibrium temperature $\Tdust$ is deduced from the balance between
the emitted and absorbed radiation from the Inter-Stellar Radiation
Field (ISRF)
\begin{equation}
    \label{eq:Teq}
    \int^{\infty}_0 B_{\omega}(\omega,\Tdust) Q(\omega)\ud\omega =
    \int^{\infty}_0 I^{ISRF}_{\omega} Q(\omega)\ud\omega
\end{equation}

The Mie theory, using the Maxwell equations of the electromagnetic
theory, leads to an exact solution for the absorption and scattering
processes by an homogeneous spherical particle of radius a whose
material is characterized by its complex dielectric constant
$\epsilon(\omega) = \epsilon'(\omega)+i\epsilon''(\omega)$.  In that
case, the absorption coefficient $Q(\omega)$ writes as an infinite
series of terms related to the complex dielectric constant.  For
particles small compared with the wavelength such as
$|\sqrt{\epsilon(\omega)}|\cdot 2\pi a/\lambda \ll 1$ (where
$\lambda=2\pi c/\omega$ is the wavelength and $c$ the speed of light
in vacuum), the absorption coefficient can be approximated by

\begin{equation}
   \label{eq:Q_Mie}
   Q(\omega) = \frac{8\pi a}{\lambda} \cdot {\rm Im}
   \left[\frac{\epsilon(\omega)-1}{\epsilon(\omega)+2}\right]
\end{equation}

In the FIR/mm wavelength range, for nonconducting materials of
astronomical dust particles , the condition $\epsilon''\ll\epsilon'$
is generally satisfied, and in this case the absorption coefficient
reduces to
\begin{equation}
   \label{eq:Q-eps}
\frac{Q(\omega)}{a}=\frac{\omega}{c} \frac{12}{(\epsilon'+2)^{2}}\epsilon'',
\end{equation}

As a consequence, following Eq.\,(\ref{eq:Q-eps}) the dust opacity rewrites
\begin{equation}
    \label{eq:k-eps}
    \kappa(\omega) = \frac{\omega}{c} \frac{9}{(\epsilon'+2)^2} \frac{\epsilon''}{\rho}.
\end{equation}

\subsection{Assumption of a $\lambda$-independent emissivity spectral index}
\label{sec:lambdaindep} 

In the absence of specific knowledge about realistic dependencies
for $\epsilon(\omega)$, a simple power law approximation for FIR/mm
dust emission is often assumed
\begin{equation}
    \label{eq:Q-beta}
    \kappa(\omega) = \kappa(\omega_0) \left( \frac{\omega}{\omega_0} \right)^\beta,
\end{equation}
where $\beta$ is referred to as the emissivity spectral index.


The intensity of thermal emission from interstellar dust, assuming
an optically thin medium, at a given wavelength
$\lambda=2\pi/\omega$ located in the FIR/mm range is

\begin{equation}
    \label{eq:I-beta}
    I_{\omega}(\omega, T)=\epsilon_{\rm e}(\lambda_0)\cdot
    \left(\frac{\omega}{\omega_0}\right)^\beta\cdot B_{\omega}(\omega,\Tdust)
\end{equation}
where $\epsilon_{\rm e}(\lambda_0)$ is the dust emissivity at a
reference wavelength 
$\lambda_0=2\pi c/\omega_0$.

Simple semi-classical models of absorption such as the Lorentz model for
damping oscillators and the Drude model for free charge carriers
(Sect.\,\ref{sec:semiclassical}) provide an asymptotic value $\beta=2$ when
$\omega \rightarrow 0$.  Such a value of the spectral index was in
satisfactory agreement with the earliest observations of the FIR/mm
SED of interstellar dust emission, but not necessarily with the most
recent ones.



The emissivity spectral index $\beta$ is equal to the slope of a
plot of the dust opacity versus wavelength in logarithmic scale.  In
the general case, the power-law assumption is not valid, and the
spectral index has to be considered as a wavelength and temperature
dependent parameter.  Similarly, $\epsilon_{\rm e}(\lambda_0)$ in
Eq.\,(\ref{eq:I-beta}) should be considered a function of~$T$.

\subsection{Relations between absorption properties of bulk material and dust}
\label{s:Relations} 

The absorption coefficient defined as the optical depth in the bulk
material per unit length is

\begin{equation}
\label{eq:alpha}
\alpha =2\mathrm{Im}(k),
\end{equation}
where $k$ is the complex value of the wave number
\mbox{$k=\omega\sqrt{\epsilon}/c$} and $\epsilon$ is the complex
dielectric constant 
Therefore for $\epsilon'' \ll \epsilon'$,

\begin{equation}
    \label{eq:alfa-omega}
   \alpha= \frac{\omega\epsilon'' }{c\sqrt{\epsilon'}}
\end{equation}

In dust and porous material measurements, a mass absorption
coefficient $\alpha/\rho$ is also often used.\\

In the following analysis (Sect.\,\ref{sec:semiclassical} and other)
we will link the macroscopic value $\epsilon$ to the
value of the local susceptibility $\chi^{}_o=\chi_o'+i\chi_o''$ defined on
microscopic scale.

Because small astronomical particles cannot always be considered as
a continuous medium, the electrodynamic equations of continuous
media must be applied with care. So we detail in the following the
equations used, and we discuss their range of application.

The Equations of motion describes the response of charges to a local
electric field $\vec{E_0}$, which permits to determine a dipole moment
per unit volume as $\vec{P}=\chi_0\vec{E_0}$.  It should be taken into
account that the local electric field $\vec{E_0}$ differs from an
external field $\vec{E}$ due to additional field produced by other
parts of the dielectric.  In the case of isotropic dielectric
materials, $$\vec{E_0}=\vec{E} +\frac{4\pi}{3}\vec{P}.$$
The dielectric constant $\epsilon$ links the external field $\vec{E}$
and the electric inductance $\vec{D}=\vec{E}+4\pi\vec{P}$ through
$\vec{D}=\epsilon\vec{E}$.  In this case, $\epsilon$ can be derived
from the susceptibility $\chi_0$, using the so called Clausius-Mossotti
equation

\begin{equation}
\label{eq:epsilon-chi}
\epsilon  = 1 + \frac{4\pi\chi_0}{1-\frac{4}{3}\pi\chi_0}.
\end{equation}
When $\epsilon'' \ll \epsilon'$, we have

\begin{equation}
\label{eq:epsilon-approxim}
\epsilon'' \approx \frac{(\epsilon'+2)^2}{9}\;4\pi\chi_o'',
\end{equation}
which leads to

\begin{eqnarray}
\label{eq:Aabs} \alpha(\omega)&=&
\frac{4\pi\omega}{c\sqrt{\epsilon'}}\frac{(\epsilon'+2)^2}{9}\;\chi_o''(\omega),\\
\label{eq:k-chi}
\kappa(\omega)&=&\frac{\omega}{c}  \frac{4\pi\chi_o''}{\rho}.
\end{eqnarray}
In Eq.\,(\ref{eq:Aabs}), the term $(\epsilon'+2)^2/9$  is the local
field correction factor, the term $\sqrt{\epsilon'}$  is the
refraction index.

Thus, the relation between the absorption coefficient $\alpha$
deduced from transmission measurements in the laboratory and the
absorption efficiency $Q$ or opacity $\kappa$, for small spherical
grains of radius $a$ and density $\rho$ can be obtained from
equations Eq.\,\ref{eq:Q-eps} and \ref{eq:alfa-omega}, and writes

\begin{equation}
    \label{eq:Q-alpha-relation}
    \frac{Q(\omega)}{\alpha(\omega)}=\frac{12a\cdot\sqrt{\epsilon'}}{(\epsilon'+2)^2}
\end{equation}
and as a consequence
\begin{equation}
    \label{eq:kappa-alpha-relation}
    \frac{\kappa(\omega)}{\alpha(\omega)}=\frac{9\cdot\sqrt{\epsilon'}}{\rho\cdot(\epsilon'+2)^2}
\end{equation}\

The condition $\epsilon''\ll\epsilon'$ is valid for dust materials
in the FIR/mm wavelength range, implying that the real part of the
dielectric constant $\epsilon'$ can be considered to first order as
a constant, leading to $d\epsilon'/d\omega=0$. This comes from the
coupling between the real part $\epsilon'$ and the imaginary part
$\epsilon''$ through the Kramers-Kronig relations exposed in
Sect.\,\ref{Sec:methods_lab}. This implies that, in that spectral
range, the absorption coefficient $\alpha$, the absorption
efficiency Q and the opacity $\kappa$ are simply proportional.  Thus
the slopes of {$\alpha(\omega)$, {Q$(\omega)$ and $\kappa(\omega)$
are the same. As for the emissivity, an absorptivity spectral index
can then be defined by the slope of the spectral plot of the
absorption $\alpha(\omega)$ in logarithmic scale.  This absorptivity
spectral index deduced from laboratory transmission measurements can
be directly compared to the emissivity spectral index deduced from
astrophysical measurements.

\subsection{The semi-classical models}
\label{sec:semiclassical} 

The well known semi-classical model of light interaction with
charge carriers in matter (electrons  and  ions) considers
the motion of charged particles driven by the time-dependent
electric field of a monochromatic plane wave, magnetic forces
being neglected compared with electrical forces.

The equations of motion of interacting charge carriers are:

\begin{eqnarray}
\label{eq:semi-clas}
&& m_j\ddot{u}_j^{\,\xi} + \sum_{j'\xi'}K_{jj'}^{\xi\xi'} u_{j'}^{\,\xi'} +
\sum_{j'\xi'}b_{jj'}^{\xi\xi'} \dot{u}_{j'}^{\,\xi'} = q_j E^{\xi}(t),
\end{eqnarray}
where $m_j$, $q_j$ are masses and charges of particles,
$K_{jj'}^{\xi\xi'}$ are constants of elastic interaction, $E^{\xi}$ is
\mbox{$\xi$-th} component of electrical field vector $\vec{E},\
u_j^{\xi}$ are components of the 3-dimensional vector of $j$-th ion
displacement.  Damping parameters $b_j$ describes
dissipation of energy.  Force constants $K$ are the coefficients of power
expansion of the potential energy $U$.

\begin{eqnarray}
\label{eq:U-vib}
&& U=\frac{1}{2}\sum_{j'\xi'}K_{jj'}^{\xi\xi'} u_{j}^{\xi}
u_{j'}^{\xi'}+...
\end{eqnarray}

$U(\cdots u_j^{\xi} \cdots)$ 
depends only on relative deviations $(u_j^{\xi}-u_{j'}^{\xi})$ with
corresponding symmetry of $K_{jj'}^{\xi\xi'}$, particularly

\begin{eqnarray}
\label{eq:Ksymmetry}
&& K_{jj'}^{\xi\xi'}=K_{j'j}^{\xi'\xi},\ 
\sum_{j'\xi'}K_{jj'}^{\xi\xi'}=0.\ 
\end{eqnarray}

Linear Eq.\,(\ref{eq:semi-clas}) with  $\vec{E}=0$ have solutions in
the form of a sum of damping oscillations

\begin{equation}
\vec{u}= \sum a_\kappa\vec{e}_{\kappa}\cdot
\exp(-\gamma_{\kappa}t+i\omega_{\kappa}t),
\end{equation}
where

\mbox{$\vec{u}=(\cdots u_{\kappa j}^{\ \xi}\cdots)$} is
the multidimensional deviation vector, $a_\kappa$ are constants,
\mbox{$\vec{e}_{\kappa}=(\cdots e_{\kappa j}^{\ \xi}\cdots)$} are
eigenvectors, and $\gamma_{\kappa}-i\omega_{\kappa}$ are eigenvalues
of matrix

\begin{equation}
\|-m_j\delta_{jj'}\delta^{\xi\xi'} + K_{jj'}^{\xi\xi'} + i\ b_{jj'}^{\xi\xi'} \|.
\end{equation}

A solution of Eq.\,(\ref{eq:semi-clas}) with
\mbox{$\vec{E}(t)=\hat{\vec{E}}e^{-i\omega t}$} is
$\vec{u}=\hat{\vec{u}}e^{-i\omega t},$
\begin{equation}
\hat{\vec{u}}=\sum_{\kappa\ j}
\frac{q_j \hat{\vec{E}}\cdot\vec{e}_{\kappa j}^*\ \omega_{\kappa}^2}
{\omega_{\kappa}^2-\omega^2-i\gamma_{\kappa}\omega}.
\end{equation}

Dielectric properties of a substance can be described by a
susceptibility $\chi_0$, which is the dipole moment per unit volume
generated by a unity local strenght $|\hat{\vec{E}}|=1$ electrical
field,

\begin{equation}
\label{eq:Lo-ro}
\chi_o = \sum_{\kappa}
\frac{\rho_{\kappa}\omega_{\kappa}^2}
{\omega_{\kappa}^2-\omega^2-i\gamma_{\kappa}\omega}.
\end{equation}
For isotropic materials, constants
\begin{equation}
\rho_{\kappa}=\Big|\sum_{j\ j'}
q_{j'}q_j\vec{e}_{\kappa j'}(\hat{\vec{E}}\cdot\vec{e}_{\kappa
j}^*)\Big|
\end{equation}
does not depend upon the direction of $\hat{\vec{E}}$.

In the extreme case where $\omega_{\kappa}\gg \gamma_{\kappa}$
the spectrum of the dielectric absorption can be
described as a collection of non-overlapping resonant {\em Lorentz} profiles

\begin{equation}
\chi_o'' = \frac{\rho_{\kappa}\omega_{\kappa}^2 \gamma_{\kappa}\omega}
     {(\omega_{\kappa}^2-\omega^2)^2+(\gamma_{\kappa}\omega)^2}.
\end{equation}

It is the case of infrared absorption bands, especially in ionic
crystals ("reststralen" bands). This corresponds to the third term
of Eq.\,(\ref{eq:semi-clas}) being small.

In the opposite case ($\omega_{\kappa}\ll \gamma_{\kappa}$)
this leads to a {\em Debye} relaxation spectrum when
$\omega\ll\omega_{\kappa}$ :

\begin{equation}\label{eq:Debye}
\chi_o'' = \frac{\rho_\kappa\ (\omega\tau_{\kappa})}{1+(\omega\tau_{\kappa})^2},
\end{equation}
where $\tau_{\kappa}=\gamma_{\kappa}/\omega_{\kappa}^2$ is a relaxation
constant. This is usually the case of polar dielectric liquids. It
corresponds to the first term of Eq.\,(\ref{eq:semi-clas}) being
negligible. 

This spectrum corresponds to situations where the inertial effects
(first term in Eq.\,\ref{eq:semi-clas}) are comparatively small.  If
this condition is satisfied, the time constant in
Eq.\,(\ref{eq:Debye}) can usually be determined by measuring or
calculating the timescale of energy dissipation during return to
equilibrium state, without solving Eq.\,(\ref{eq:semi-clas}).

A third extreme case, corresponding to the second term of
Eq.\,(\ref{eq:Debye}) being negligible is usually refered to as the {\em
Drude} approximation.  It applies to free carriers and is considered in
Sect.\,\ref{drudemodel}.

\subsection{The Lorentz model for crystalline insulators}
\label{LorentzModel}

The Lorentz model applies for crystalline isolators with an ionic
character, and describes the resonant interaction of an
electromagnetic wave with the vibrations of ions which act like
bound charges. Such an interaction is caused by the electric dipoles
which arise when positive and negative ions move in opposite
direction in their local electric field.  This kind of dust
absorption can be clearly identified in astrophysical observations
as emission or absorption bands.  Indeed, the optical depth $\tau$
is proportional to 
opacity $\kappa$ (Eq.\,\ref{eq:k-chi}):

\begin{equation}
    \label{eq:Lo-2} 
     \kappa(\omega)
     = \frac{4\pi}{c\rho} \sum_{\kappa}
     \frac{\rho_{\kappa}\omega_{\kappa}^2 \gamma_{\kappa}\omega^2}
     {(\omega_{\kappa}^2-\omega^2)^2+(\gamma_{\kappa}\omega)^2}
\end{equation}

The maximum number of optically active vibrational modes in a crystal
is $n_{opt}=3(s-1)$, where $s$ is the number of atoms in an elementary
cell of the lattice.  They are usually classified as branches in
dispersion dependence of oscillation frequency $\omega(k)$ on wave
number $k$.  All these branches are restricted by relatively high
$\omega(k)$ in the infrared region.  Another 3 branches with
$\omega(k)\rightarrow 0$ when $k \rightarrow$0 are acoustic modes.
They are optically not active in crystal because a total charge of cell
is zero.  Another situation for amorphous matter is considered here
in Sect.\,\ref{sec:mex}.

For substances of astrophysical interest, the characteristic
frequencies of bands fall in region \mbox{$\lambda \la 50\mic$}.
Therefore, in the FIR/mm wavelength range, the asymptotic $\omega
\rightarrow 0$ of Eq.\,(\ref{eq:Lo-2}) can be used:

\begin{equation}
    \label{eq:Lo-0}
    \kappa(\omega) =
    \omega^2\ \frac{4\pi}{c\rho} \sum_{\kappa}
    \frac{\rho_{\kappa}\gamma_{\kappa}}{\omega_{\kappa}^2}
\end{equation}

Thus, the Lorentz model formally applied to dust in the FIR/mm
wavelength range gives a spectral index $\beta=2$, a value often used
in "standard" dust models.

Many dust models have related the long wavelength absorption by
interstellar silicates and ices ($\lambda>100\mic$) to the damping
wing of the infrared-active fundamental vibrational bands, visible
at shorter waves (see for example Andriesse et al.
\cite{Andriesse74} and references therein).

Laboratory infrared data are satisfactorily described (see for
example Rast \cite{Rast68}) by a sum of resonant Lorentz profiles
(or reststrahlen bands) as described by Eq.\,(\ref{eq:Lo-ro}) and
Eq.\,(\ref{eq:Lo-2}), in which the various constants $\rho_{\kappa}$,
$\omega_{\kappa}$, $\gamma_{\kappa}$ are determined from
spectroscopic measurements.  This model gives a satisfactory
approximation for the infrared region, not only for ionic crystals,
but also for covalent ones and for their amorphous counterparts with
an appropriate choice of $\rho_{\kappa}$, $\omega_{\kappa}$,
$\gamma_{\kappa}$.

The situation is more complicated for the low frequency region,
where absorption is defined by damping constant $\gamma$, in
contrast to IR bands, where absorption is weakly dependent on
$\gamma$. The damping constants $b_j$ introduced in the
semi-classical model (Eq.\,\ref{eq:semi-clas}) and therefore
$\gamma_{\kappa}$ in Eq.\,(\ref{eq:Lo-2}) have no simple microscopic
explanation. The damping is linked to heating or distribution of
energy to all modes of grain vibrations.  So $\gamma$ characterizes
intermode interactions which are frequency and temperature dependent
in general. In contrast to single phonon absorption of resonant IR
photons, a wing absorption is a two phonon difference process.  Due
to the conservation of energy, the energy of the absorbed low
frequency photon is equal to the energy difference of the two
phonons in the fundamental vibrational bands (Rubens and Hertz
\cite{Rubens12}, Ewald \cite{Ewald22}). In fact in crystals, any
combination of acoustical and
 optical phonons which satisfies the energy and momentum
 conservation rules can take part to the absorption . This is the main source
 of infrared absorption in homoatomic crystals. In the FIR wavelength range
  the two-phonon difference processes are responsible for
 the absorption in ionic crystals. Because the efficiency of such processes are
 entirely determined by the difference between the phonon occupation numbers
of the two involved phonons, the corresponding absorption decreases
drastically at decreasing temperatures (Hadni \cite{Hadni70}).

However the optical behavior of amorphous materials is different.
  The disorder of the atomic arrangement leads to
a breakdown of the selection rules for the wavenumber. The
absorption spectra reflects thus the whole density of vibrational
states. This induces in the longest wavelength range a broad
absorption band due to single phonon processes which completely
dominates over the multiphonon absorption (H. Henning and H.
Mutschke \cite{Henning97}). As a consequence, such lattice
absorption does not vanish at low temperature for the amorphous
solids , in contrast to the crystalline ones.

\subsection{The Drude model for conducting materials}
\label{drudemodel}

For free charge carriers the previous semi-classical equation of motion
(Eq.\,\ref{eq:semi-clas}) can still be used, setting all restoring
force coefficients $K$ to zero, equaling all the masses $m_j$
to the effective mass of the free carriers $m_c$ (
$m_j=m_c$), and using a constant value for the damping
parameters \mbox{$b_j/m_j$=const=$\gamma_c$}.
Therefore Eq.\,(\ref{eq:Lo-ro}) for the susceptibility simplifies as
\mbox{$\chi_o =  -\ \omega_p^2\ (\omega^2-i\gamma_c\omega)^{-1}$},
where $\omega_p$ is the so-called plasma frequency
\mbox{$\omega_p^2 = N_c q_c^2/m_c$},
and $N_c$ is the number of free carriers per unit volume.

In metals, 
the number of free electrons per unit volume is independent of
temperature, and is of the same order of magnitude than the number
of atoms (\mbox{$N_c \sim \ 10^{22}$\,cm$^{-3}$}).
This leads to a frequency $\omega_p\sim 10^{15}$\,s$^{-1}$ located in
ultraviolet or in the visible, and to correspondingly high conductivity
$\sigma=\omega_p^2/\gamma_c$.

In the long wavelength range, far from the plasma frequency
($\omega\ll\omega_p$) and when the modulus of $\epsilon$ is large
compare to 1, the asymptotic behavior of the opacity is given by

\begin{equation}
    \label{eq:Drude-asymptotic}
    \kappa(\omega) = \omega^2 \frac{9\gamma_c^2}{c\rho\ \omega_p^4},
\end{equation}
which also gives a spectral index $\beta=2$.
Metallic materials and high conductivity semiconductors  are expected
to follow this model.
Possible dust absorbents of such type are graphite and metallic iron.

Therefore, both the Lorentz and the Drude models in the low
frequency limit yield $\beta$=2 in Eq.\,(\ref{eq:Q-beta}).  These
very different models have common features.  Their characteristic
times (\mbox{$\omega_{\kappa}^{-1}$}, $\omega_p^{-1}$) are far from
FIR/mm region.  In this case a good approximation is obtained by
asymptotic expansion of the absorption coefficient $\alpha(\omega)$
on even powers of frequency $\omega^{2n}$, $n>0$.  In the general
case the first term is dominating ($n=1$) and then we have
$\beta$=2.

Other values (e.g. $n$=2, $\beta$=4) are possible if
physical mechanisms suppressing the $n=1$ term are present, as in the case of
screening considered in Sect.\,\ref{sec:mex}.  Intermediate values of
$\beta$ provide a clear evidence for specific processes which time
constants are of the order of $\omega^{-1}$.

\section{Temperature and spectral variations of dust optical properties}
\label{Sec:Evidences}

\subsection{Astronomical evidences}
\label{s:astro}

Analyses of millimeter and submillimeter emission from molecular
clouds have found spectral indices between $\beta$=1.5 (Walker, Adams
\& Lada,1990) and $\beta$=2 (Gordon 1988,Wright et al., 1992).
However some values in excess of 2 can also be found in the
literature (Schwartz 1982, Gordon 1990). A few studies of dense
clouds have yielded spectral indices around 1 (Gordon 1988, Woody
et al. 1989, Oldham et al. 1994), but  observations of disks of
gas and dust around young stars  can indicate values sometimes
less than 1 (Chandler et al. 1995). A submillimeter continuum
study of the Oph 1 cloud (Andre, Ward-Thompson \& Barsony 1993)
found a typical value of $\beta$ equal to 1.5 but attributed changes
in observed ratios to temperature variations. Without appropriate
temperature data, the authors could not be conclusive on this
issue.

More recent observations have evidenced that the actual spectral energy
distribution of dust emission could be significantly more complicated
than described above.  Analysis of the FIRAS results have shown that,
along the galactic plane, the emission spectrum is significantly
flatter than expected, with a slope roughly compatible with $\beta=1$
(see Reach et al. \cite{Reach95}, in particular their Fig.~7).  They attributed the
flattening compared to the $\beta=2$ canonical value to an additional
component peaking in the millimeter range and favored that the
millimeter excess could be due to the existence of very cold dust at
$\Tdust=5-7\,K$ in our Galaxy.  However, the origin of such very cold
dust remained unexplained, and the poor angular resolution of the FIRAS
data ($\simeq 7\degr$) did not allow further investigations.

Finkbeiner et al.  \cite{Finkbeiner99} later adopted a similar model.
They showed that a good fit of the overall FIRAS data could be obtained
using a mixture of warm dust at $\Tdust\simeq16\,K$ and very cold dust
at $\Tdust \simeq 9\,K$.  The temperature of the warm component was set
using a fit to the DIRBE data near $\lambda=200\mic$ while the
temperature of the very-cold component was set assuming an independent
dust component immersed in the same radiation field but with different
absorption properties.
They implied that these two components could be graphite (warm) and
silicates (very cold). However, they gave no quantitative argument
to support this hypothesis

PRONAOS was a french balloon-borne experiment designed to determine
both the FIR dust emissivity spectral index and temperature, by
measuring the dust emission in four broad spectral bands centered at
$200, 260, 360$ and $580\mic$ (see Lamarre et al.\cite{Lamarre94},
Serra at al.\cite{Serra02},  Pajot et al.\cite{Pajot06}). The
analysis of the PRONAOS observations towards several regions of the
sky ranging from diffuse molecular clouds to star forming regions
such as Orion or M17 have evidenced an anti-correlation between the
dust temperature and the emissivity spectral index (see Dupac et al.
\cite{Dupac03}). The dust emissivity spectral index varies smoothly
from high values ($\beta\simeq2.5 $) for cold dust at $\Tdust\simeq
12\,K$ to much lower values ($\beta=1$) for dust at higher
temperature (up to $\Tdust\simeq 80\,K$) in star forming regions
like Orion. Dupac et al. (\cite{Dupac03}) showed that these
variations were not caused by the fit procedure used and were
unlikely to be due to the presence of very cold dust.  They
concluded that these variations may be intrinsic properties of the
dust.

Extragalactic observations have also revealed unexpected behavior
of dust emission at long wavelength.  For instance, Galliano et
al. (\cite{Galliano03}, \cite{Galliano04}),
combining JCMT data at $450$ and $850\mic$, IRAM 30m data at 1.2
mm with ISO and IRAS measurements in the IR and FIR evidenced a
strong millimeter excess towards a set of four blue compact galaxy
(NGC1569, II Zw40, He~2-10 and NGC 1140).  They attributed this
excess to the presence of very cold dust ($\Tdust=5-7\,K$) with a
very flat emissivity index ($\beta=1$) which would have to be
concentrating into very dense clumps spread allover the galaxy.
This very cold dust would then account for about half or more of the total
dust mass of those galaxies.


\subsection{Laboratory evidences}
  \label{Sec:methods_lab}

\subsubsection{Comparing laboratory and astronomical data}

As it was shown in Sect.\,\ref{s:Relations}, laboratory data on
absorption spectral index could be used for interpreting
astronomical observations, taking into account the quantitative
limitations considered here.  Some other limitations should be
considered.

First, there is a difference between bulk laboratory samples and the
small cosmic dust.  The latter probably has a high surface to volume
ratio, and surface effects are important for spectroscopy as for the
physics and the chemistry of dust.

Second, samples used for laboratory measurements are generally
synthesized in very small amounts. Along with the small values of the
absorption coefficients it requires high-sensitivity dedicated
instruments such as $^3$He-cooled bolometers or/and high power sources
of radiation.

In the laboratory indirect IR/FIR/mm spectroscopy methods are also used,
such as Kramers-Kronig spectroscopy.  The dielectric constant
counterparts $(\epsilon'-1)$ and $\epsilon''$ are Hilbert transforms of
each other and are related through the Kramers-Kronig relations (see
Landau \& Lifshits \cite{Landau82}) by

\begin{eqnarray}
\epsilon'(\omega) &=& \nonumber
    \frac{1}{\pi}\int^{\infty}_{-\infty}\frac{\epsilon''(x)}{x-\omega}\ \ud x +1,
\\ &&\label{eq:KK}
\\ \epsilon''(\omega) &=& \nonumber
-\frac{1}{\pi}\int^{\infty}_{-\infty}\frac{\epsilon'(x)}{x-\omega}\ \ud x,
\end{eqnarray}
where both integrals are the Cauchy principal values.

A simple form of Eq.\,(\ref{eq:KK}) is for $\omega=0$:

\begin{equation}
\label{eq:KK0}
\epsilon'(0)=\frac{2}{\pi}\int^{\infty}_0 \epsilon''(x)\ \ud\ln x+1.
\end{equation}

The use of these relations together with reflection measurements
\mbox{$R(\omega)=|(\sqrt{\epsilon}-1)/(\sqrt{\epsilon}+1)|^2$}
allows to resolve the system of equation with respect to
$\epsilon'(\omega)$ and $\epsilon''(\omega)$ and therefore to
calculate an absorption spectrum.  Possible method inaccuracies due
to inverse problem solution should be considered carefully.  In
particular, it is sometimes assumed as a prior that the solution has
the form of a power spectrum.  In practice, the most reliable
approach should be to use the same model for interpretation of both
the astrophysical and laboratory data.

\subsubsection{Spectral index variations in disordered solids}
\label{sec:Sch}

It has been known for more than 25 years, that disordered
solids (glasses and mixed crystals) have significantly higher submillimeter
absorption than perfect crystals of similar chemical nature.

Mon et al. (\cite{Mon75}) performed first absorption measurements on
various amorphous dielectric materials in the millimeter wavelength
range (1\,mm - 5\,mm) and at low temperature (in the range 0.5\,K -
10\,K). A strong temperature-dependence of the absorption was
observed, leading to an millimeter absorption excess at  low
temperature. They already attributed this effect to the presence of
a distribution of Two-level Systems in the studied materials.

Others experimental results on bulk materials at frequencies
$\omega/2\pi c$ between 0.1 and 150\,cm$^{-1}$ and $T$ between 10\,K
and 300\,K were discussed by Strom \& Taylor (\cite{Strom77}).  The
spectra follow a general empirical relation $\alpha = K\omega^2$
between 15 and 150\,cm$^{-1}$, where the temperature independent
constant $K$ depends on the material considered and its internal
structure.  At lower frequencies $\omega/2\pi<$15\,cm$^{-1}$
($\lambda > 700\mic$) the slope of the spectra is variable ($\beta$
between 0 and 3) and temperature dependent as shown in Strom \&
Taylor (\cite{Strom77}).

B\"osch (\cite{Bosch78}) performed absorption measurements for
temperature between 300\,K and 1.6\,K in the FIR/mm range
(500$\mic$ - 10\,mm) on amorphous silicates mainly composed of
SiO$_2$.  The studied silicate is not a typical grain analog.
However, measurements performed over such a wide spectral and
temperature range are particularly appropriated to study physics
governing absorption processes.  B\"osch found a strong temperature and
frequency dependency of the absorption coefficient in the millimeter
range ($\lambda>500\mic$).  The absorption coefficient was
characterized by an increase of the spectral index with temperature
from $\beta=1.6$ at 300\,K to $\beta=3$ at 10\,K and by a
strong temperature dependence.  To describe this temperature and
frequency behavior, B\"{o}sch also referred to the existence of Two Level
Systems in the material.

Therefore, laboratory spectroscopy shows evidences for deviation of
absorption spectral index from the canonical $\beta=2$ in the FIR/mm.
The temperature independent absorption process which provides 
$\beta=2$ in the spectral region $\lambda<700\mic$ does not seem to
dominate the absorption at longer wavelengths.  We will see in Sect.
\ref{Sec:ModelPhysics} that the model proposed here includes this
phenomenon in a natural manner.

\subsubsection{Spectral variations in dust analogues} 
\label{Labevidences}

FIR/mm optical properties of solids with chemical composition and
structure considered as analogues of interstellar dust were investigated
in a few laboratory studies (for example Koike et al.  \cite{Koike80},
Agladze \cite{Agladze96}), Henning and Mutschke \cite{Henning97},
Mennella et al. \cite{Mennella98}, Boudet et al.  \cite{Boudet05}).
Particular interest has been given to low temperature investigations of
amorphous or other disordered materials, which reveal systematic
deviations from the phonon theory of crystal solids and was interpreted
using the 
TLS theory (see B\"osch \cite{Bosch78},
Phillips \cite{Phillips87} and K\"uhn \cite{Kuhn01}).

A few laboratory measurements have been performed on amorphous solids
in the FIR/mm range at low temperature.  However, some of these studies
were performed over a wide range of temperature and revealed a
variation of the spectral behavior of the absorption coefficient with
temperature.

Agladze et al.  (\cite{Agladze96}) performed absorption measurements on
typical interstellar analog grains (crystalline and amorphous silicate
grains) at low temperature (1.2 - 30\,K) in the millimeter range
(700\,$\mic$- 2.9\,mm).  They found an unusual behavior of the
millimeter absorption of amorphous silicates : the absorption
coefficient decreases with temperature down to about 20\,K and then
increases again with decreasing temperature.  They described the
frequency dependence of the absorption coefficient using a
temperature-dependent spectral index.  Depending on the samples, the
spectral index can decrease from $\beta=2.6$ at 10\,K down to $\beta=1.8$
near 30\,K. They referred to tunneling effect in Two Levels Systems to
explain this behavior.

Mennella et al.  (\cite{Mennella98}) 
measured the temperature dependence
of the absorption coefficients between 295\,K and 24\,K and
for wavelength between 20$\mic$ and 2\,mm on different kind
of cosmic grain analogs (amorphous carbon grains, amorphous and
crystalline silicates ).  They reported a significant
temperature-dependence of the spectral index of the absorption
coefficient, particularly strong for their amorphous iron-silicate
sample.  Their measurements showed a systematic decrease of the
spectral index with increasing temperature.

Boudet et al.  (\cite{Boudet05}) performed measurements on different
types of amorphous silicates (typical analog grains and SiO$_2$
samples) for temperature between $\rm 10\,K$ and $\rm 300\,K$ and
wavelength from $\rm 100\mic$ to $\rm 2\,mm$. They found a strong
temperature and frequency dependency of the absorption coefficient.
They defined two spectral index depending on the wavelength range
considered.  For wavelengths between $\rm 500\mic$ and $\rm 1\,mm$
they found a pronounced decrease of the spectral index with
increasing temperature whereas, for wavelengths between $100\mic$
and $200\mic$, the spectral index was found to be constant with
temperature. They put some SiO$_2$ sample through a strong thermal
treatment to remove most SiOH groups, and observed that the
temperature-dependent absorption disappears totally. They thus
identified the silanol groups as the defects that, in their silicate
sample, are at the origin of the behavior. Considering that the OH
groups can not simply increase the coupling between the photon and
the bulk phonons, they assumed reasonable that the defects induce
closely spaced local energy minima, which correspond to the dynamics
of a "particle" in an asymetric double-well potential.

These laboratory studies provide strong indications that the
absorption coefficients of amorphous grain analog materials vary
substantially both with temperature and frequency.


\section{A model to explain temperature and spectral variation of
 dust optical properties}
\label{Sec:ModelPhysics}

\subsection{Modeling disordered structure}

Two mechanisms have to be considered when dealing with amorphous
materials, in order to take disorder into account.

First, one should consider the acoustic oscillation excitation based
upon the interaction of solids with electromagnetic waves due to
Disordered Charge Distribution (See Schl\"omann
\cite{Schlomann64}).  This effect takes place not only in amorphous
media, but also in disordered crystals like mixed and polycrystals, and
in some monocrystals with inverse spinel structure, for example.  This
absorption is temperature independent.  The DCD model introduces a
correlation length $l_c$ which quantifies the scale over which the
atomic structure of the material realizes charge neutrality.  The
absorption spectrum of such a structure presents two asymptotic
behaviors.  Towards short wavelengths, the absorption is characterized
by a spectral index $\beta=2$ and in the longest wavelength range, the
spectral index tends towards $\beta=4$.  The location of the spectral
range in which the transition between the two regimes is directly
related to the correlation length.  The DCD model is described in
Sect.\,\ref{sec:mex}.

The second mechanism describes the disorder at atomic scale, as a
distribution of tunneling states,
which leads to a
temperature-dependent optical absorption, and enables to explain the
temperature dependence of the spectral index observed in laboratory
experiments.  This model was originally elaborated to explain the
unusual thermal and optical properties of amorphous material at low
temperatures and has been described in details by Phillips
(\cite{Phillips72}), Anderson, Halperin \& Varma (\cite{Anderson72}) (see
also Phillips \cite{Phillips87} for a review).  In particular, the TLS
model was developed to explain the fact that, at low temperatures, the
specific heat of amorphous solids exceeds what is expected from the
Debye theory for solids.  This anomalous behavior is common to all
amorphous materials and therefore appears independent of their detailed
chemical composition and structure.  The existence of such two levels
systems has been pointed out by means of microwave and submillimeter
spectroscopy experiments (e.g. Agladze et al.  \cite{Agladze96},
B\"osch \cite{Bosch78} and reference therein).  The TLS model is
described in Sect.\,\ref{Sec:TLS}.

The two mechanisms considered here for the absorption by interstellar
dust probably dominate in the FIR/mm and longer wavelengths domain.
Both have a large degree of universality without specific signatures
characterizing the chemical nature of the absorbing substance.  On the
other hand, they are sensitive to the internal structure of the solids,
in particular their degree of ordering, and to
mechanical properties such as density and elasticity.

\subsection{DCD absorption}
\label{sec:mex} 

\subsubsection{DCD Theory}

In amorphous materials, the lack of long-range order permits single
phonon/photon interactions with excitation of any modes of
mechanical vibrations.  Low frequency vibrations are not linked to
molecular vibrational bands like is the case for ice or silicate
bands.  In infinite media they correspond to traveling acoustic
waves.  In finite
bodies like interstellar dust grains, these modes can be described as 
standing waves.  So we use the term "phonon" here for quantum of
vibrational motion not restricting it (if not stated otherwise) to
periodic lattice or infinite media.

The concept of phonon quasiparticles (Tamm, \cite{Tamm30}) arises as
the result of quantization of vibrational motion in the harmonic
approximation, which takes into account only quadratic terms (on atom
displacements) of energy in Eq.\,(\ref{eq:U-vib}).  It corresponds to
Eq.\,(\ref{eq:semi-clas}) with $b_j=0$.  Dissipation effects arise in the
phonon concept as the result of phonon interaction, taking into account
anharmonic terms in Eq.\,(\ref{eq:U-vib}).  Their role was pointed first
by Debye (\cite{Debye14}).

For harmonic oscillations $u =\hat{u}e^{-i\omega t}$ with electrical
field $\vec{E}(t)=\hat{\vec{E}}e^{-i\omega t}$
Eq.\,(\ref{eq:semi-clas}) takes the form

\begin{equation}
\label{eq:u-omega} 
-\omega^2\ \hat{u}_j^{\xi} +\sum_{j'\xi'} D_{jj'}^{\xi\xi'}\ \hat{u}_{j'}^{\xi'} = 
\frac{q_j}{m_j}\ \hat{E}^{\xi},
\end{equation}
where the coefficients of the dynamics matrix are\\
$D_{jj'}^{\xi\xi'}=K_{jj'}^{\xi\xi'}/m_j$. \\

The solution of the dispersion equation
\begin{equation}
\label{eq:dis-D}
-\omega^2\ e_j^{\xi}+D_{jj'}^{\xi\xi'}\ e_{j'}^{\xi'}=0
\end{equation}
leads to resonant frequencies $\omega_{\kappa}$ and corresponding
eigenvectors $\vec{e}_{\kappa}=(\cdots e_{\kappa j}^{\ \xi}\cdots)$.
In an orthonormal basis of vectors $\vec{e}_{\kappa}$, when
\begin{equation}
\label{eq:norma}
\sum_{\xi\ j} e_{\kappa j}^{\xi} e_{\kappa'j}^{*\xi}=\delta_{\kappa\kappa'},
\end{equation}
the equation~(\ref{eq:u-omega}) has a simple form
\begin{equation}
\label{eq:u-norm} 
(\omega_{\kappa}^2-\omega^2 )\ \hat{u}_{\kappa} =
\vec{g}_{\kappa}\cdot\vec{\hat{E}},
\end{equation}
where $\hat{u}_{\kappa}$ are amplitudes of normal
oscillations,
\begin{displaymath}
g_{\kappa}^{\xi} = \sum_{j} q_j m_j^{-1} e_{\kappa j}^{*\xi},
\end{displaymath}
and $^{*}$ denotes complex conjugates.

The susceptibility corresponding to Eq.\,(\ref{eq:u-norm}) is
given by the tensor

\begin{equation}
\chi^{\xi\xi'}_0=\sum_{\kappa j'j}\frac {q_{j'} q_j m_j^{-1}
e_{\kappa j}^{*\xi} e_{\kappa j'}^{\xi'}}
{\omega_{\kappa}^2-\omega^2-i\omega\gamma_{\kappa}},
\end{equation}
or in the macroscopically isotropic case by the scalar

\begin{displaymath}
\chi_0=(\chi^{xx}_0+\chi^{yy}_0+\chi^{zz}_0)/3
\end{displaymath}
and therefore,

\begin{equation}
\chi_0''={\rm Im}\ \chi_0=\frac{1}{3}\sum_{\kappa j'j\xi} \frac
{\omega\gamma_{\kappa} q_{j'} q_j m_j^{-1} e_{\kappa j}^{*\xi}
e_{\kappa j'}^{\xi}}
{(\omega_{\kappa}^2-\omega^2)^2+(\omega\gamma_{\kappa})^2}.
\label{eq:chi-res}
\end{equation}

Damping factors $\gamma_{\kappa}$
were introduced here to get a stationary solution. $\gamma_{\kappa}$
values depends on phonon interactions, as discussed above.
In the FIR/mm region, overlapping of resonance curves produces a quasi continuous
absorption spectrum.
The integrated spectrum in Eq.\,(\ref{eq:chi-res})
can be simplified replacing the integral of the sum of weak individual
profiles by the sum of their integrated values.

\begin{equation}
\int_{-\infty}^{+\infty}{\rm Im} \frac{\ud\omega}
{\omega_{\kappa}^2-\omega^2-i\omega\gamma_{\kappa}}
 = \frac{\pi/2}{\sqrt{\omega_\kappa^2-\gamma_\kappa^2}}\ .
\end{equation}

\begin{equation}
\label{eq:chi-avg}
\left<\ \chi_0''(\omega)\ \right> =\frac{\pi}{6\omega}
\sum_{|\omega-\omega_{\kappa}|<\Delta\omega/2} \frac{q_{j'} q_j
m_j^{-1} e_{\kappa j}^{*\xi} e_{\kappa j'}^{\xi}}{\Delta\omega},
\end{equation}
where $\Delta\omega$ is an averaging interval satifying

\begin{equation}
\omega\gg\Delta\omega\gg\gamma_\kappa.\label{eq:cond-S}
\end{equation}

A number of states is proportional to $\Delta\omega$, therefore the
result does not depend on $\Delta\omega$ and $\gamma_{\kappa}$
when condition in Eq.\,(\ref{eq:cond-S}) is satisfied, which therefore also
defines the validity region of the results of Vinogradov
(\cite{Vinogradov60}) and Schl\"omann (\cite{Schlomann64}).

In macroscopically uniform and isotropic cases the eigenvectors of
Eq.\,(\ref{eq:dis-D}) in the form of plane waves are

\begin{equation}
\label{eq:plane-w}
e_{\kappa j}^{\xi}=
\tilde{e}_{\vec{k}\eta}^{\;\xi}\exp(i\vec{k}\cdot\vec{r}_j).
\end{equation}

The manyfold of solutions of the dispersion equation is ordered here in
accordance with vector $|\vec{k}|$
and splits into several branches
labeled here by index  {\small $\eta$}. Each branch has a different
dispersion curve $\omega=\omega_\eta(|\vec{k}|)$. Most branches are
restricted to high frequencies (infrared) and only three branches
extend down to the FIR/mm region. They are so called "acoustical" modes.
For such low frequencies, these have a linear dispersion dependance

\begin{equation}\label{eq:cond-acoustic}
\omega_\eta=\upsilon_\eta\ k,
\end{equation}
where $k=|\vec{k}|$, and $\ \upsilon_\eta$ is speed of sound.  One branch
corresponds to longitudinal propagation ({\small $\eta=l$}), the two
others to 2
polarizations of transverse propagation ({\small $\eta=t$}).
Numericaly, $\upsilon_t<\upsilon_l$.  Displacements of neighbor ions in sound oscillations
are synchronous and $\tilde{e}_{\vec{k}\eta}^{\;\xi}$ does not depend on
$j$.  The vectors of displacements $\vec{e}_{\vec{k}\eta}=
[\tilde{e}_{\vec{k}\eta}^{\ x},\tilde{e}_{\vec{k}\eta}^{\ y},\tilde{e}_{\vec{k}\eta}^{\ z}]$
are oriented such that
$\vec{e}_{\vec{k}l}\parallel\vec{k},\ \vec{e}_{\vec{k}t}\perp\vec{k}$.
Their absolute values are defined by the normalization equation
Eq.\,(\ref{eq:norma}).  The number of normal oscillation per unit volume in
the frequency interval $\Delta\omega$ is $N_\omega\Delta\omega$, where

\begin{equation}
N_\omega=\frac{\omega^2}{4\pi^{2}\upsilon_\eta^3}
\end{equation}
and therefore

\begin{equation}
\label{eq:q2m}
\chi_0''(\omega)=\frac{\omega N}{24\pi}
\left(\frac{2}{\upsilon_t^3}\left < \frac{q^2}{m} \right >_{\omega/\upsilon_t}+
\frac{1}{\upsilon_l^3}\left < \frac{q^2}{m} \right >_{\omega/\upsilon_l}\right),
\end{equation}
where $N$ is number of ions per unit volume and

\begin{equation}
\label{eq:q2msp}
\left < \frac{q^2}{m} \right >_k = \frac{1}{N^2}
\sum_{j,j'=1}^N \frac{q_j q_{j'}}{m_j}\exp(i\vec{k}\cdot(\vec{r}_j-\vec{r}_{j'}) ).
\end{equation}

The  equation derived here is a generalization of the Schl\"omann's
(\cite{Schlomann64}) expression for one type of disordered ions in a
perfect crystal \mbox{$\langle q^2\rangle _k^{}m^{-1}$}, which was defined
by a spatial power spectrum of charge distribution
$\langle q^2\rangle_k$ only.

\subsubsection{Spectrum of DCD absorption}

The absorption spectrum corresponding to Eq.\,(\ref{eq:q2msp})
can be calculated using correlation functions. 
It is a cross-correlation function for general case
\begin{equation}
\phi_c(\Delta\vec{r})=
\Big< \frac{q_j q_{j'}}{m_j}\Big > \big(\ \overline{q^2_{\rm m}}\ \big)^{-1}
\end{equation}
Here $\langle\ \rangle$ designate an ensemble average,
\mbox{$\Delta\vec{r}=\vec{r}_j-\vec{r}_{j'}$}.  Parameters
$\overline{q_1^2},\ \overline{q^2_{\rm m}}$ are described below and do not
depend on $\Delta\vec{r}$.

The correlation function can be determined from a physical model of
considered material.  The first, simplest model has an uncorrelated charge
distribution and therefore a delta-function correlation

\begin{equation}\label{eq:delta-corr}
\phi(\Delta\vec{r})=\delta(\Delta\vec{r})=
\delta(\Delta x)\delta(\Delta y)\delta(\Delta z).
\end{equation}

In general case a cross-correlation function, Eq.\,(\ref{eq:delta-corr})
gives a flat spatial spectrum with intensity not depending on $k$:

\begin{equation}
\Big < \frac{q^2}{m} \Big >_k = \overline{q^2_{\rm m}},
\end{equation}
and therefore

\begin{equation}
\label{eq:chi-delta}
\chi_0''(\omega)=\frac{\omega}{12\pi \upsilon_t^3}\ \overline{q^2_{\rm m}}
\end{equation}
Here, we omitted the second term in
Eq.\,(\ref{eq:q2m}) since $\upsilon_l^{-3}\ll 2 \upsilon_t^{-3}$.

An autocorrelation function should be used for $m$=const
\begin{equation}
\phi_a(\Delta\vec{r})=
\langle\ q(\vec{r}_j) q(\vec{r}_i)\ \rangle\ (\ \overline{q_1^2}\ )^{-1},
\end{equation}
with a flat spatial spectrum $\langle q^2\rangle_k = \overline{q_1^2}$
for $\delta$ correlation (Eq.\,\ref{eq:delta-corr}) leading to Eq.\,(\ref{eq:chi-delta})
with a constant $\overline{q^2_{\rm m}}=\overline{q^2_1}/m$.

In perfect crystals the charge distribution is entirely correlated for
long distances (long range order) and the absorption mechanism
described by Eq.\,(\ref{eq:q2m}) and Eq.\,(\ref{eq:chi-delta}) does not
take place.  Nevertheless even for crystals with a perfect lattice, a
disordered charge distribution is possible if the lattice permits a non
unique configuration of the charges in an elementary cell, as is the
case, for example of the cubic lattice of salt (NaCl).  In such cases,
only stochastically distributed charges should be included in the
calculation of
$\overline{q^2_1}$, which can then be defined as

\begin{equation}
\label{eq:q1}
\overline{q^2_1} =\frac{1}{N_1}\sum_{j=1}^{N_1} (\Delta q_j)^2
< \frac{1}{N_1}\sum_{j=1}^{N_1} q_j^2,
\end{equation}
where summing is to be done only for ions involved in the disordered
distribution and $\Delta q$ denotes deviations from the mean charge value,
which is different for different nodes of the crystal lattice.
The ordered part of the charge distribution is linked only to the spectral components
with \mbox{$\omega\ \widetilde{>}\ \upsilon_{\eta}/d$}, where $d$ is the lattice
period.  This frequency range corresponds to mid infrared absorption
bands and are not considered here.

In amorphous media the corresponding coefficient $\overline{q^2_{\rm
m}}$ should also exclude the regular part of the charge distribution
produced by the short range order in the medium.  An equation similar
to Eq.\,(\ref{eq:q1}) cannot be written in the general case.  The effect
of this short range screening can be expressed by the inequality 

\begin{displaymath}
\overline{q^2_{\rm m}} < \frac{1}{N}\sum_{j=1}^{N} \frac{q_j^2}{m_j},
\end{displaymath}

The second model of Schl\"omann (\cite{Schlomann64}) can be interpreted as 
screening over large scales. Large-scale electrical neutrality of crystals
and glasses could be a result of
high charge mobility in solutions or melts from which the corresponding
materials were manufactured.

The corresponding correlation function should be written in generalized case as
\begin{equation}
\phi(\Delta\vec{r})=\delta(\Delta\vec{r})-\frac{1}{8\pi l_c^3}\exp(|\Delta\vec{r}|/\lcor),
\end{equation}
where $\lcor$ defines the correlation length.
The corresponding spectra are given by

\begin{equation}
\langle q^2\rangle_k = \overline{q^2_1}\ g(k\lcor),
\label{eq:gk-w}
\end{equation}
where

\begin{equation}
\label{eq:g-x}
g(x) = 1-(1+x^2)^{-2}
\end{equation}
and

\begin{equation}
\label{eq:wo-lc}
\chi_0''(\omega)=
\frac{\omega}{12\pi\upsilon_t^3}\ \overline{q^2_{\rm m}}\ g(\omega/\omega_c),
\end{equation}
where

\begin{displaymath}
\omega_c=\upsilon_t/\lcor.
\end{displaymath}

The spectral dependence of Eq.\,(\ref{eq:g-x}) is

\begin{equation}
\label{eq:2-4}
    g\Big(\frac{\omega}{\omega_c}\Big) \left\{
    \begin{array}{ll}
        \sim\omega^2/\omega^2_c &\textrm{for }\omega\ll\omega_c\\
        \approx 1 &\textrm{for }\omega\gg\omega_c
    \end{array} \right.
\end{equation}

This is also true of other often used correlation function shapes,
such as step correlation functions :

\begin{equation}
    \phi(\vec{r})= \left\{
    \begin{array}{ll}
        \delta(\vec{r})-3/(4\pi l_c^3)&\textrm{for }|\vec{r}|<l_c\ ,\\
        0 &\textrm{for }|\vec{r}|\ge l_c\ .
    \end{array} \right.
\end{equation}

The spectral dependence in Eq.\,(\ref{eq:2-4}) is common for charge
distributions, which are uncorrelated over short distances and respect
neutrality over larger scales.

The DCD absorption is temperature independent.  The corresponding
absorption coefficient $\alpha(\omega)\sim \omega \chi_{o}''$
(Eq.\,\ref{eq:Aabs}) presents two asymptotics behaviors on both sides
of $\omega_c$ which depends on the correlation length.  In the high
frequency range, the spectral index takes the value \mbox{$\beta=2$},
and for the lower frequency range, the spectral index is equal to
\mbox{$\beta= 4$}.  This frequency dependence absorption is shown on
Fig.~\ref{fig:km0} for various values of $l_{c}$.

\begin{table*}[ht]
\caption[]{\label{tab:par}Parameters of disordered solids.}
\begin{flushleft}
\begin{tabular}{ l c c c c c c c c c } 
\hline
\hline
Material & $\rho$ & $\epsilon$ & $\upsilon_t$ & $l_c$ & $\langle q^2\rangle/e^2$ &$\mu_{\rm b}^{ }$
& $\overline{P}\mu_{\rm b}^2$ & $\gamma_{\rm e}$ & Refs. \\
& g/cm$^3$ & & cm/s  & nm & & D & & eV & \\
\hline
Silica Glass (SiO$_2$)& 2.2 & 3.8 & $4.1\E{5}$ & 1-4 & 1.1 & 0.18$^a$ &
$1\E{-5}$ & 1.5 & 1,2,3 \\
Soda-silica glass$^b$ (Na$_2$O$\cdot$3SiO$_2$) & 2.9 & 2.6 & $3\E{5}$ & 3 & 1 &&
$1.4\E{-3}$ & 0.5 & 4\\
Chalogenic glasses & & & & 1-4 & 0.1-1 & 0.5-5 & & & 2 \\
Inverse spinel crystals & & & $4\E{5}$ & & 1/14 & & & & 5\\
\hline
\hline
\end{tabular}
\end{flushleft}
\begin{list}{}{} 
\item[$^{\mathrm{a}}$] A previously reported value of dipole momentum 0.6 D was
overestimated due to the absence of local field correction.
\item[$^{\mathrm{b}}$] Barrier heights distribution parameters:
maximum at $V_{\rm m}/\kB=550\,K$, width $V_0/\kB=410\,K$, $V_{\min}/\kB=50\,K$.
\item[References:] 1: Hubbard et al. \cite{Hubbard03},
2: Strom and Taylor \cite{{Strom77}}, 
3: Phillips \cite{Phillips87}, 4: B\"osch \cite{Bosch78}, 5: Schl\"omann \cite{Schlomann64}.
\end{list}
\end{table*}

\begin{figure}
\includegraphics[width=8cm]{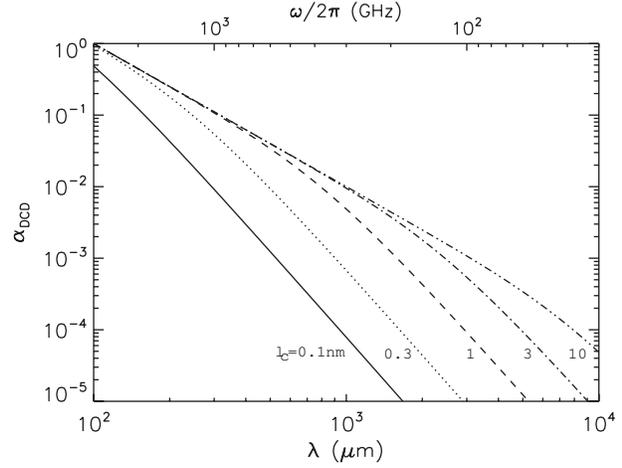}
\caption{\label{fig:km0}
Absorption spectrum due to a Disordered Charge Distribution.
The absorption coefficient $\alphaDCD$
is given in arbitrary units, normalized at
$\lambda=100\mic$ and $\lcor=10\,nm$.
The solid, dot, dash, dash-dot and dash-dot-dot lines correspond
to $\lcor=0.1$, 0.3, 1, 3 and $10\,nm$ respectively, assuming a
transverse sound velocity \mbox{$\upsilon_t=3\E5$\,cm/s.}}
\end{figure}

\subsection{TLS Absorption}
\label{Sec:TLS}

\subsubsection{Phenomenological TLS Theory}

At low temperatures, the thermal and dielectric properties of
disordered solids (e.g. glasses and mixed crystals) show definite
deviations from the predictions of the phonon theory developed for
perfectly ordered crystals.  Most of these phenomena can be described
in the formalism of the so called Two Level Systems (TLS), which is the simplest model of
tunneling states.  In comparison to DCD it could be considered as a
second approximation for the description of electromagnetic properties
of solids linked to disordering.  Vacancies unavoidable in disordered
lattices produce in first approximation a possibility for chaotic
distribution of some charges, which was considered in the previous
section.  This distribution is considered as frozen in the lattice due
to the large height of barrier dividing vacancies in comparison with
the available thermal energy.  The TLS theory takes into account
transitions of charges due to quantum tunneling possible at any
temperature.

The TLS theory is the result of a macroscopic analysis of phenomena as
opposed to an exact microscopic description.  It is based on the
hypothesis of a flat distribution of tunneling states with energy level
differences sufficiently low with respect to the usual vibrational
energy.  The temperature dependance of two-level system properties
results from deviations from the model of harmonic oscillator, which is
a good approximation for the description of crystal lattice vibrations.
The origin of TLS is vacancies in the lattice which gives the
possibility, for atoms or groups of atoms, to have different
equilibrium spatial configurations.  As a result, the potential curve
has at least a double-well form (DWP, see Fig.~\ref{f:asim}) in
contrast to single-well potential (SWP) of harmonic oscillators in
crystal without vacancies.

\begin{figure}
  \begin{center}
    \includegraphics[width=7cm]{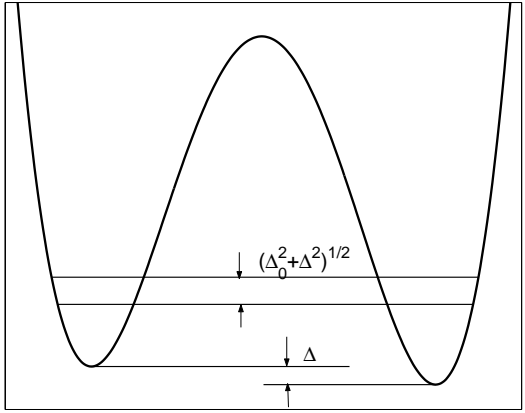}
  \end{center}
\caption{
\label{f:asim}
Asymmetric double-well potential. Ordinate is the potential energy.
Abscissa is a coordinate in configuration space, in the simple case
a rotation angle or a linear displacement of groups of atoms
in the lattice.  $\Delta$ is the value of potential asymmetry.  The
difference in energies of the lowest levels results from the asymmetry
$\Delta$ and tunnel splitting $\Delta_0$.
}
\end{figure}

In the TLS model described e.g. by (Phillips \cite{Phillips87}), the
energy distribution of states is defined under the assumption that the
asymmetry $\Delta$ of DWP and the height $V$ of the DWP potential
barrier are stochastic values and their probability distribution can be
assumed constant over a certain range of $V$ and $\Delta$ values.  Like
for the well known tunneling probability, the energy splitting
$\Delta_0$ has an exponential dependence with barrier width $d$:

\begin{equation}
\label{eq:delta0}
\Delta_0=\hbar\Omega\exp[-d(2mV)^{1/2}/\hbar],
\end{equation}
where $\hbar\Omega=(E_1+E_2)/2$ is the average energy of levels and $m$ is
the effective mass.  This exponential dependence
along with the flat probability distributions of $\Delta$ and $V,$ leads to
the 2-dimensional distribution

\begin{equation}
\label{eq:P-Delta}
P(\Delta_0,\Delta)=\overline{P}/\Delta_0,
\end{equation}
where $\overline{P}$ is a constant. The integral of $P$ over $\Delta_0$
does not diverge because the lowest $\Delta_0$ values are limited by
the highest energy $V$ through Eq.(\ref{eq:delta0}).

The exact form of $P(\Delta_0)$ in the vicinity of $\Delta_0^{\min}$ is
not important in most cases because the corresponding energy
($\Delta_0^{\min}/\kB\ll 0.01\,K$) are very far from values under
consideration.  
A value of constant $\overline{P}$ is not defined by
the TLS theory, it varies from one material to another, and can be
regarded as a free parameter to be determined from experiments or
observations.

In oder to better understand the TLS {\em phenomenological} approach,
K\"uhn (\cite{Kuhn01}) built a {\em microscopic} model using the translational
invariant Hamiltonian

\begin{eqnarray}
\label{eq:Hamilton}
H &=& \sum_j\frac{p_j^2}{2m}+U(\{u_j\}),
\nonumber \\ \label{eq:Kuhn}
U &=& \frac{1}{4}\sum_{ij}J_{ij}(u_i-u_j)^2+\frac{g}{2N}\sum_{ij}(u_i-u_j)^4,
\end{eqnarray}
where $p_j$ are impulses, $u_j$ are displacements of $j$-th atom
from some reference points. The first term is a quadratic potential
with random coefficients $J_{ij}$. The second term is a stabilizing quartic potential (not random)
with a constant $g>0$ coefficient.
The stochastically simulated coefficients $J_{ij}$ lead to disordered
equilibrium arrangement of atoms.  Disordered potential relief
(Eq.\,\ref{eq:Kuhn}) reproduces an ensemble of states both SWP and DWP,
with a broad spectrum of barrier heights $V$ and asymmetries $\Delta$.

Some TLS effects associated with excited-state transitions of TLS
could be brought to the fore in some laboratory experiments
(FitzGerald et al.\cite{FitzGerald94} and Sievers et al.
\cite{Sievers98}) on a number of mixed fluorite crystals and
soda-lime silica glass. However these effects can only be seen in a
narrow temperature range, roughly below 15K, before relaxation
processes dominate at higher temperature. These effects will be
incorporated to our model in a forthcoming paper. In the following,
we restrict our consideration to the two lowest levels with energy
difference $E=E_2-E_1=\sqrt{\Delta^{2} +\Delta_0^{2}}$.

As opposed to thermal properties, the TLS absorption spectrum depends
not only on the TLS density of states distribution $P$ over energies
$\Delta_0$, $\Delta$, but also on the values of the matrix elements of
the dipole transitions between levels

\begin{equation}
\langle 1 | \mu | 2 \rangle=\mu_{\rm b}\frac{\Delta_0}{E},
\end{equation}
where $\mu$ is an electric dipole moment,

\begin{equation}
\mu_{\rm b}=\frac{1}{2}\frac{\partial \Delta}{\partial \xi},
\end{equation}
where $\xi$ is the value of the electrical field.

The dipole moment $\mu_{\rm b}$ is the second TLS parameter, which
characterizes
its interaction with the electromagnetic field.  For isotropic
materials the
square of dipole moment should be averaged over all directions as
$\langle \mu_{\rm b}^2 \rangle=\mu_{\rm b}^2/3$. Like the parameter $\overline{P}$
discussed above, the parameter $\mu_{\rm b}$ varies from one material to
another, and can be regarded as a free parameter, to be determined by
experiments or observations.  In some experimental publications, values of
the dipole moment are given, which are not corrected for the local field and
orientation averaging, $\mu'=\mu_{\rm b}(\epsilon'+2)/3\sqrt{3}$.

The TLS absorption spectrum shape can be obtained by solving the Bloch
equations which describe the interaction of TLS with electromagnetic
wave and lattice oscillations.  The latter interaction is described by
the third material dependent parameter of TLS theory, i.e. the elastic dipole
moment given by :

\begin{equation}
\gamma_{\rm e}=\frac{1}{2}\frac{\partial\Delta}{\partial e},
\end{equation}
where $e$ is the value of the strain field.  The order of magnitude of
$\mu_{\rm b}$ and $\gamma_{\rm e}$ is about 1\,D and 1\,eV, respectively.

The solution of general equations for the TLS absorption can be described as 
a result of two different contributions:
an absorption which has a resonant character and an attenuation which
has the typical form of a relaxation.  In practice it is convenient to
treat these processes separately.  B\"osch \cite{Bosch78} explained his
experimental results considering three processes:
resonant tunneling , relaxation due to phonon-assisted tunneling,
relaxation due to phonon-assisted hoping over the potential barrier.
In the next sections, we will discuss each of them in more details.

\subsubsection{Resonant absorption}\label{s:res} 

The resonant absorption of a photon of energy $\hbar \omega$ concerns
only those TLS characterized by the splitting energy $E_{2}-E_{1} =
\hbar \omega$ .  Let $N(\omega)$ be the density per unit frequency
of those TLS, and $N_{1}(\omega)$ and $N_2(\omega)$ the densities
per unit frequency of those TLS which are in the state of energy $E_1$
and $E_2$ respectively.  Of course these densities are strongly
temperature dependent, but in the TLS model only the
two lower levels are considered and it is assumed that $N=N_1 + N_2$.

The resonant tunneling absorption then takes the form

\begin{equation}
\alpha_{\rm res}=\frac{\hbar\omega}{c\sqrt{\epsilon'}}
B_{12}\Big(\frac{\epsilon'+2}{3}\Big)^2(N_1-N_2),
\end{equation}
where $B_{12}$ is the Einstein coefficient for absorption,

\begin{displaymath}
B_{12}=\frac{4\pi^2\langle 1 |\mu| 2 \rangle^2}{3\hbar^2}.
\end{displaymath}

The population of the levels under the thermal equilibrium hypothesis
 gives

\begin{displaymath}
N_1-N_2=N(\omega)\tanh(\hbar\omega/2\kB T),
\end{displaymath}
and taking into account that $E=\hbar\omega$

\begin{equation}
\label{eq:Fitz}
\alpha_{\rm res}=\frac{4\pi^2\omega}{3c\sqrt{\epsilon'}}
\Big(\frac{\epsilon'+2}{3}\Big)^2\tanh(\hbar\omega/2\kB T)\ G(\omega),
\end{equation}
where

\begin{equation}
\label{eq:G-p}
G(\omega)=\frac{\overline{P}\mu_{\rm b}^2}{(\hbar\omega)^2}\int_0^{\hbar\omega}
\frac{\pG (\Delta_0)\Delta_0\ud\Delta_0}{\sqrt{1-(\Delta_0/\hbar\omega)^2}}.
\end{equation}

Equation\,(\ref{eq:Fitz}) was derived in the work of Hubbard et al.
(\cite{Hubbard03}), the term $G(\omega)$ is the so called Optical
Density Of States (ODOS) and can be determined from laboratory
spectroscopy.

The coefficient $\pG (\Delta_0)\la 1$ we introduced in
Eq.\,(\ref{eq:G-p}) to describe deviation of $P(\Delta_0)$ from
Eq.\,(\ref{eq:P-Delta}) at large $\Delta_0$.  To avoid confusion in this
work we will not change definitions of variables where it is possible,
preferring transformation of the original equations taken from different
sources.

In the TLS theory, confirmed by low-temperature measurements of
thermal properties, \mbox{$\pG =1$} for energies smaller than
$10^{-3}$\,eV.
If \mbox{$\pG =1$} for all $Delta_0$,
Eq.\,(\ref{eq:G-p}) gives $G(\omega)=\overline{P}\mu_{\rm
b}^2$=const.  Expression for the TLS resonant tunneling absorption,
corresponding to \mbox{$\pG=1$}, was used by B\"osch
(\cite{Bosch78}) and Schickfus et al. (\cite{Schickfus75},
\cite{Schickfus76}):

\begin{equation}
\label{eq:B78R}
\alpha_{\rm res} =
\frac{4\pi^2 \overline{P}\,\mu'^2}{c\sqrt{\epsilon'}}\omega\;\tanh(\hbar\omega/2\kB T).
\end{equation}

The frequency independent ODOS was found to be in satisfactory
agreement with experiments for wavenumbers $\omega/2\pi c
<$12\,cm$^{-1}$.  Its measured values ($\overline{P}\mu_{\rm b}^2$) are
given in Table\,\ref{tab:par}.  Assuming $\mu'= 1D$, the density of
states is \mbox{$\overline{P} = 0.9\cdot 10^{33}$
erg$^{-1}$\,cm$^{-3}$} for soda-lime-silica glass.

More accurate spectroscopic measurements in wide spectral regions
show a definite drop of $G(\omega)$ to zero within the error level,
for $\omega$ higher than some frequency $\omega_{\rm m}$. Agladze \&
Sievers (\cite{Agladze98}) measured a profile with a cut-off at
20\,cm$^{-1}$ in high density amorphous phase ices of H$_2$O and
D$_2$O. Fitzgerald, Sievers \& Campbell (\cite{Fitzgerald01a})
detected a sharp cut-off in fluoride mixed crystal spectra, with
wavenumber $\omega_{\rm m}/2\pi c=$13\,cm$^{-1}$. Hubbard et al.
(\cite{Hubbard03}) observed a shallow cut-off at $\omega_{\rm
m}/2\pi c=$15\,cm$^{-1}$, 9\,cm$^{-1}$ and 6\,cm$^{-1}$ for the soda
lime silica, the SiO$_2$ and the germanium glasses, respectively.

Hubbard et al.  (\cite{Hubbard03}) suggested that the distribution
$P(\Delta_0,\Delta)$ (Eq.\,\ref{eq:P-Delta}) has a cut-off at energy
$E_{\rm m}=\hbar\omega_{\rm m}$.  In our Eq.\,(\ref{eq:G-p}) it
corresponds to $\pG(\Delta_0)$ in the form of a step function

\begin{equation}
    \label{eq:Pconst}
     \pG(\Delta_0) = \left\{
     \begin{array}{ll}
         1& \textrm{for $\Delta_0<E_{\rm m}$}\\
         0& \textrm{for $\Delta_0>E_{\rm m}$.}
     \end{array} \right.
\end{equation}

Under this assumption, Hubbard et al.  (\cite{Hubbard03}) calculated
a frequency dependent expression for the ODOS by means of integration of
distribution function equivalent to Eq.\,(\ref{eq:G-p}).
For our consideration it is convenient to write it in the form

\begin{equation}
\label{eq:G-g}
G(\omega) = g(\omega/\omega_{\rm m}) \overline{P}\mu_{\rm b}^2.
\end{equation}

The ODOS of (Hubbard et al. \cite{Hubbard03}) is described by the following
function :

\begin{equation}
    \label{eq:H03P}
    g(\omega/\omega_{\rm m}) = \left\{
       \begin{array}{ll}
       1 & \textrm{for $\omega/\omega_{\rm m}\le 1$} \\
       1-\sqrt{1-(\omega/\omega_{\rm m})^{-2}} &
       \textrm{for $\omega/\omega_{\rm m}>1$}.
       \end{array} \right.
\end{equation}

However, the spectra corresponding to the ODOS proposed by
Hubbard et al.  (\cite{Hubbard03}) (Eq.\,\ref{eq:H03P})
exhibits a prominent break which was not observed in
laboratory spectra of noncrystaline materials and has not been observed in
spectra of astronomical dust emission.

We suggest that the cut-off of the distribution function should be
complemented by a continuity condition and therefore continuity of
$\pG(\Delta_0)$.  The simplest form of continuous function for
$\pG(\Delta_0)$ is considered in Appendix A and leads to the following
a shape of the ODOS

\begin{equation}
\label{eq:ODOS}
    g(x) =  \left\{
    \begin{array}{ll}
        1+x^2 g_1(x) & \mathrm{for} \ x<1,\\
        1+x^2 g_1(x)-g_2(x)\sqrt{1-x^{-2}} & \mathrm{for} \ x>1.
    \end{array}\right.
\end{equation}
where \mbox{$x=\omega/\omega_{\rm m}$}, polynoms $g_1(x),\ g_2(x)$ are defined
in Appendix by Eq.\,(\ref{eq:g1}) and Eq.\,(\ref{eq:g2}).

This ODOS function has no break at $\omega=\omega_{\rm m}$, it is
in better agreement with laboratory measurements of amorhous materials.
It is also in full agreement with TLS theory at $\omega\ll\omega_{\rm m}$.
For this reason it  will be used in the calculations of Sect.\,\ref{s:astro-model}.

\begin{figure}
    \includegraphics[width=8cm]{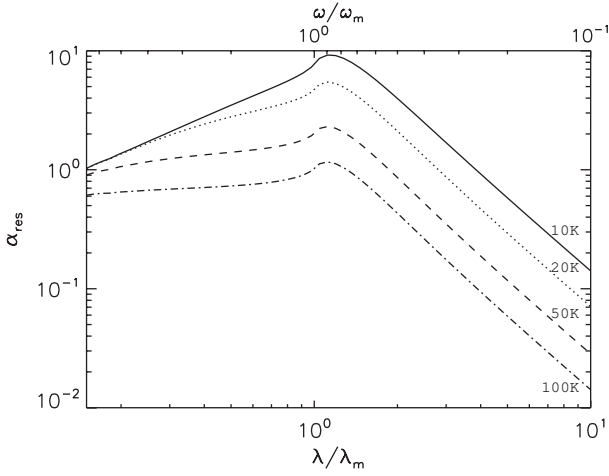}
\caption{
\label{fig:kmI}
Absorption spectrum due to resonant tunneling in Two Level Systems.
The absorption coefficient is given in relative units normalized at
$\lambda=100\mic$ and $T=10\,K$.  The abscissa axis is normalized to the
cut-off wavelength of the Optical Density of States $\lambda_{\rm m}=700\,\mic$ (see text).
The solid, dot, dash and dash-dot lines correspond to $T=10$, 20, 50
and $100\,K$ respectively.  The various curves can be scaled using the
simple relation $\lambda_{\rm m} T$=const.  In the low frequency
wings, curves scale as $\alpha\sim\omega^2/T$.
}
\end{figure}

\subsection{Relaxation processes} 
\label{s:relax}

Direct consideration of the relaxation processes permits to evaluate
the nonresonant part of TLS absorption spectra through analysis of the
mechanism of the TLS level population relaxations after some deviation
from equilibrium.  The relaxation time constant $\tau$ is determined by
the rate ($\tau^{-1}$) of transitions between levels, in which
transitions due to intensive interactions with the lattice thermal
oscillations (phonons) dominate.

The rate of the TLS transitions as a result of strain field generated by thermal
fluctuations in a lattice was evaluated by Phillips (\cite{Phillips72}).

\begin{equation}
\label{eq:tunnel}
\tau(\Delta_0,E)\ ^{-1}=\tau_1^{-1}\ \Delta_0^2\ E^{-2},
\end{equation}
where

\begin{equation}
\label{eq:tau-a}
\tau_1^{}=a\ E^{-3}\tanh(E/2\kB T)
\end{equation}
and

\begin{equation}
\label{eq:aconst}
a=\pi \upsilon_t^5 \rho \hbar^4 \gamma_{e}^{-2}
\end{equation}
is a material constant.
Parameters included in Eq.\,(\ref{eq:aconst}) were already described above.

Phillips (\cite{Phillips72}) calculated also an equilibrium value of
the TLS dipole moment $\overline{\mu}$ and therefore the resulting
susceptibility $\chi(\omega)$ in the form of a Debye spectrum (Eq.
\,\ref{eq:Debye}):

\begin{equation}
\label{eq:Debye-Phill}
\chi_0''(\omega)=\frac{\mu_{\rm b}^2}{3\kB T}\frac{\Delta^2}{E^2}
\sech^2\Big(\frac{E}{2\kB T}\Big)
\frac{\omega\tau}{1+(\omega\tau)^2}
\end{equation}

The rate of {\em tunneling relaxation} (Eq.\,\ref{eq:tunnel}) is
proportional to the tunneling probability, so it does not take into
account direct transitions in TLS which energy is higher than the
barrier height.  This transition rate is proportional to the number
of phonons with such energy, which should follow a Boltzman
distribution and depends on temperature and on the energy barrier
height $V$.  In glasses with $V/\kB \ga 500$\,K the time of {\em
hopping relaxation} $\tau$ varies with $T<100$ K in accordance with the
Arrhenius relation (Hunklinger \& Schickfus \cite{Hunklinger81}).

\begin{equation}
\label{eq:tau-hopping}
\tau=\tau_0\ \exp(V/\kB T)
\end{equation}
with $\tau_0\approx 10^{-13}$\,s (fused silica).

Experimental values of $V$ and $\tau_0$ were determined using the
temperature dependence of the position of the spectrum maximum
($\omega=\tau^{-1}$, see Eq.  \,\ref{eq:Debye-Phill}).  The
distribution of TLS on $V$ leads to the broadening of the corresponding
spectrum, sometimes preventing definite detection of its maximum.  The
measurement of the spectrum broadening permits to evaluate the width
and shape of the TLS distribution on $V$.  In practice experimentalists
often use the shape of a truncated Gaussian probability distribution to
fit the energy distribution of TLS.

\begin{equation}
\label{eq:Gauss-P}
P(V)=C_V\ \pV(V),
\end{equation}
where

\begin{equation}
    \label{eq:Gauss-trunk}
    \pV(V)= \left\{
    \begin{array}{ll}\exp(-(V-V_{\rm m})^2/V_0^2)
    &\textrm{ for }V>V_{\min},\\
        0 &\textrm{ for }V<V_{\min},
    \end{array} \right.
\end{equation}
where $C_V$ is the usual normalization coefficient (Eq.\,\ref{eq:norm})
depending on $V_{\rm m},\ V_0,\ V_{\min}$, which are
parameters of the distribution shape. An example of measured numerical values of
these parameters is given in Table\,\ref{tab:par}.

Starting from known distribution of TLS on energies $E,\ \Delta_0,\ V$
and spectrum (Eq.\,\ref{eq:Debye-Phill}) for fixed values of $E,\
\Delta_0,\ V$ one can synthesize a final relaxation spectrum.  The
integrated spectra are considered in the following sections.

\subsubsection{Tunneling relaxation} 

The absorption spectrum due to phonon-assisted tunneling relaxation
can be obtained by integration of Eq.\,(\ref{eq:Debye-Phill}) on $E$
and $\Delta_0$. The expression for this was provided by Fitzgerald,
Campbell \& Sievers.(\cite{Fitzgerald01})

\begin{equation}\label{eq:FC01}
\alphaphon = A\cdot\omega\cdot F(\omega,T),
\end{equation}

where

\begin{equation}
A=\frac{\overline{P}\mu_{\rm b}^2}{3c\sqrt{\epsilon}} \Big(
\frac{\epsilon'+2}{3} \Big)^2\label{eq:A-Fitz}
\end{equation}

is a material constant and :

\begin{eqnarray}
\label{eq:f2wT}
F(\omega,T)&=&\frac{1}{2\kB T} \nonumber
\int^{\infty}_{0}\int^{\infty}_{\tau_1^{}}\sqrt{1-\frac{\tau_1^{}}{\tau}
}\ \sech^2\Big(\frac{E}{2\kB T}\Big) \frac{\omega\ \ud\tau\ \ud
E}{1+\omega^2\tau^2},
\label{eq:f2i}
\end{eqnarray}

with $\tau_1$ given by Eq.\,(\ref{eq:tau-a}).  A change of
variables made using relation between $\Delta_0$ and $\tau$ for
integral calculation.

Another, simplified equation describing tunneling relaxation absorption was
used by Jackle (\cite{Jackle72}) and by B\"osch (\cite{Bosch78})
\begin{eqnarray}\label{eq:B78}
    \alphaphon &=& A \int^{\infty}_{0}
    \frac{\ud E}{2\kB T} \sech^2\Big(\frac{E}{2\kB T}\Big)
    \frac{\omega^2\tau_1}{1+\omega^2\tau_1^2},
\end{eqnarray}
where $\tau_1$ is given by Eq.\,(\ref{eq:tau-a}),
parameters $A$ and $a$ (in Eq.\,\ref{eq:tau-a}) are to be defined from
experiments.
The use of Eq.\,(\ref{eq:B78})
is equivalent to evaluating the integral in Eq.\,(\ref{eq:Debye-Phill})
substituting values by that for a symmetric DWP $\Delta$=0.
Formally it is equivalent to replacing $\sqrt{1-\tau_1/\tau}$ by $\tau_1\delta(\tau-\tau_1)$.
Our evaluation shows large inaccuracy in this procedure for
values of 
absorption. For details see Appendix~\ref{ap:tun}.

However, this approach gives satisfactory accuracy for time constant evaluation
linked to parameter $a$.
For soda-silica glass (B\"osch \cite{Bosch78})
$a=4.2\E{-56}$ erg$^3$s.  It corresponds to $\gamma_{e}=0.5$\,eV
(see Table\,\ref{tab:par}).

\begin{figure}
    \includegraphics[width=8cm]{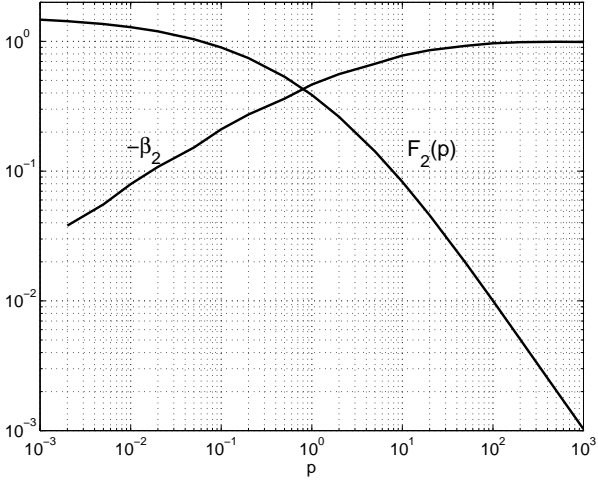} \vspace{3mm}
    \caption{\label{fig:f2}
The function $\Fphon(p)$ (solid line) and the exponent of its slope
\mbox{$\beta_2=d\ln \Fphon/d\ln p$} (dotted line) plotted as a
function of the parameter $p=\aphon\omega/(2\kB T)^3$ (see text).
The spectral index of phonon absorption $\betaphon$ is related to
$\beta_2$ as $\beta=1+\beta_2$ and ranges from $\betaphon=1$ at low
frequencies (low p values) to $\betaphon=2$ at high frequencies
(high p values).  The temperature dependence $p(T)$ generates a
shift of the absorption spectra along the frequency axis. }
\end{figure}

\begin{figure}
    \includegraphics[width=8cm]{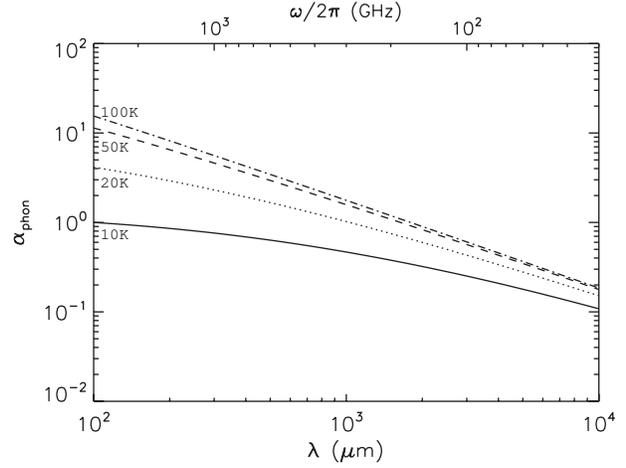} \vspace{3mm}
\caption{
 \label{fig:phon} Spectra of absorption due to tunneling
relaxation $\alpha_{\rm phon}$ in the FIR/mm.  The absorption
coefficient is given in arbitrary units, normalized at
$\lambda=100\mic$ and $T=10\,K$.  Line styles are the same as in
Fig.\,\ref{fig:kmI} }
\end{figure}

The full equation (\ref{eq:f2wT}) for calculation of phonon relaxation spectra
can be simplified without loss of accuracy. We made it using a
function of one  argument $\Fphon(p)$ instead of a function of two arguments
$F(\omega,T)$ in Eq.\,(\ref{eq:FC01}).
We introduce the intermediate parameter $p(\omega,T) = a\omega/(2\kB T)^3$.
This permits to use a one-dimensional grid of argument,
efficiently reducing computing time. The use of interpolation
in precalculated values (Table\,\ref{tab:f2} in Appendix)
make use of this function as fast as a standard one.

Therefore an accurate and relatively simple equation for tunneling relaxation spectrum is
\begin{equation}\label{eq:t-r}
\alphaphon = A\cdot\omega\cdot \Fphon (a\omega(2\kB T)^{-3}),
\end{equation}
where $A$ (Eq.\,\ref{eq:A-Fitz}) is a material constant and the function
$\Fphon$ is given by
\begin{eqnarray}\label{eq:F2}
\Fphon (p)&=&\int^1_0\int^{\pi/2}_{x_1} \sqrt{ 1-\frac{\tan x_1}{\tan x} }\ \ud x \ud y,
\\x_1 &=&\arctan( p\ y\textrm{ arcth}^{-3} y ).
\end{eqnarray}
A plot of function $\Fphon (p)$ is shown on Fig.~\ref{fig:f2}.

The absorption spectrum for tunneling relaxation is temperature
dependent in accordance with Eq.\,(\ref{eq:FC01}). As opposed to
resonant absorption, the tunneling relaxation absorption increases
with temperature, as shown on Fig.~\ref{fig:phon}

\subsubsection{Hopping relaxation}\label{s:hopping}

The expression for the absorption spectrum due to hopping relaxation
can be obtained by integrating Phillip's equation
(\ref{eq:Debye-Phill}) over a Gaussian distribution of barrier height
$\pV(V)$ given in Eq.\,(\ref{eq:Gauss-trunk}) :

\begin{equation}\label{eq:hopping}
\alpha_{\rm hop} =
B_{\rm hop}(T)\ \omega\ \int^{\infty}_{0} \frac{\omega\tau}{1+\omega^2\tau^2} P(V)\ \ud V ,
\end{equation}
where $\tau=\tau_0\exp(V/\kB T)$ and
the coefficient $B_{\rm hop}(T)$ does not depend on $\omega$.

Examples of spectra described by Eq.\,(\ref{eq:hopping}) are shown on
Fig.~\ref{fig:hop1Hz}.  It is known from general principles that the
asymptotic behavior of $\epsilon(\omega)$ when \mbox{$\omega\rightarrow 0$}
leads to \mbox{$\alpha(\omega)\sim\omega^2$} (spectral
index $\beta$=2).  Figure~\ref{fig:hop1Hz} shows that this region
begins at $f<10$\,MHz for $T$= 100\,K and at $f<100$\,Hz for $T<
50\,K$.

\begin{figure}
\includegraphics[width=8cm]{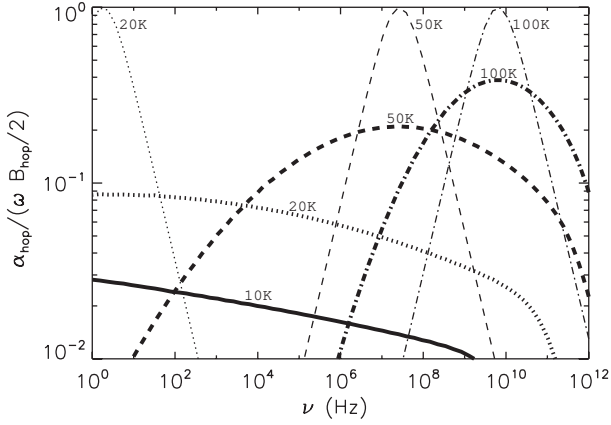}
\caption{
\label{fig:hop1Hz}
Hopping relaxation absorption spectra in the frequency range from 1\,Hz to
$1$\,THz.  The ratio of the absorption coefficient $\alpha_{\rm
hop}$ to the parameter $\omega B_{\rm hop}/2$ is plotted to show the
whole spectrum at comparable scales and to evidence variations of the
peak position with temperature.  The curves are computed using a
barrier height of $V_{\rm m}/\kB$=550\,K (see Tab.\,\ref{tab:par}).
Line styles are the same as in Fig.\,\ref{fig:kmI}.  Thin curves
correspond to unbroadened spectra with $V_0\ll V_{\rm m}$.  Thick
curves correspond to $V_0/\kB=$410\,K.
}
\end{figure}

\begin{figure}
\includegraphics[width=8cm]{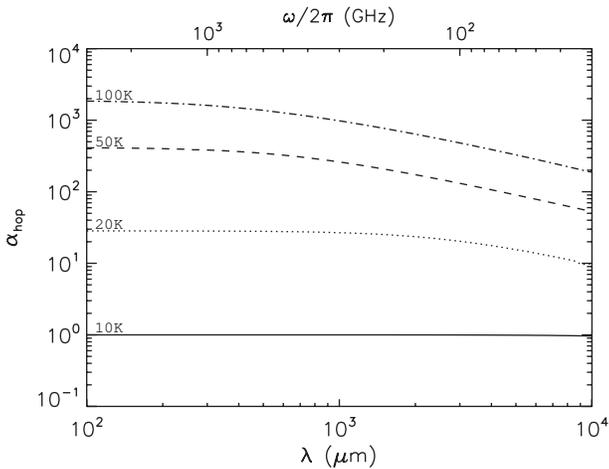}
\caption{
\label{fig:hop-mm}
Spectra of absorption due to hopping relaxation $\alpha_{\rm hop}$ in
the FIR/mm. The absorption coefficient is given in arbitrary units,
normalized at $\lambda=100\mic$ and $T=10\,K$.  Line styles are the
same as in Fig.\,\ref{fig:kmI} In this relatively narrow spectral
region, the behavior of the curves is in agreement with the
approximation of Eq.\,(\ref{eq:hop-mm}).
}
\end{figure}

At high frequencies (beginning from 
mm-wave region) the spectra shown on Fig.~\ref{fig:hop-mm}
are determined by the low energy part of the barrier height distribution.
It can be approximated by 
\begin{equation}
\alpha_{\rm hop} \approx B_{\rm hop}(T)\omega  P_{\min}
\int^{\infty}_{V_{\min}} \frac{\omega\tau}{1+\omega^2\tau^2}\ \ud V,
\end{equation}
where $P_{\min}=P(V_{\min})$, and therefore
\begin{equation}\label{eq:hop-mm}
\alpha_{\rm hop} \approx B_{\rm hop}(T)\omega P_{\min}\kB T\ \arctan
\Big(\frac{e^{-V_{\min}/\kB T}}{\omega\tau_0}\Big).
\end{equation}
For spectra on Fig.~\ref{fig:hop-mm},
a value of $P_{\min}$ corresponds to 
$P_{\min}\,\kB=$\mbox{$\exp[-(V_{\min}-V_{\rm m})^2/V_0^2]/V_0\sqrt{\pi}$}
= 3.11$\E{-4}$\,K$^{-1}$.

B\"osch (\cite{Bosch78}) used an approximated expression 
$B_{\rm hop}=h_r/T$, where $h_r$ is a constant.
The approximation used by Fitzgerald et al.  (\cite{Fitzgerald01}) gives $B_{\rm
hop}=2\pi\mu^2/3c\sqrt{\epsilon'}$ which is a material constant.
%
In Annex \ref{a:hopping}, we derive a more accurate approximation which gives
\begin{equation}\label{eq:hopping-log}
B_{\rm hop}(T)=\frac{8\pi}{3c\sqrt{\epsilon'}}\frac{(\epsilon'+2)^2}{9}
\overline{P}\mu_{\rm b}^2(\cD+\ln T),
\end{equation}
where T is in Kelvin and material constant $\cD$ is given by
\begin{equation}
\cD=\ln\frac{\kB}{\Delta_0^{\min}}+\ln 4-1+\int_0^1\ln{\rm arcth}\ x\ \ud x
\end{equation}
in accordance with Eq.\,(\ref{eq:hopping-log}) and
(\ref{eq:hopping-C}). $\cD$ is therefore an additional parameter of
tunneling states. Its value corresponds to  tunnel splitting in low
$\Delta$ region where tunneling state density distribution becomes
significantly less than that of Eq.\,(\ref{eq:P-Delta}).
The TLS theory was proved to apply in low temperature experiments at
$T$ down to 2 mK (Phillips \cite{Phillips87}), therefore
$\Delta_0^{\min}/\kB<2\E{-3}$\,K and $\cD$ should verify $\cD\ga
5.8$. The dependence of $B_{\rm hop}(T)$ is relatively weak and a
tenfold variation in $T$ from 10\,K to 100\,K leads to variation of
$B_{\rm hop}(T)$ less than $30\%$.

A difference in temperature dependence of considered approximations is significant,
especially for absorption calculations over a wide range of temperatures. It does
not manifest itself when fitting laboratory data over a limited region of temperature,
leading only to a biased value of a parameter $A$. 
The main temperature dependence of $\alpha_{\rm hop} $
is defined by exponential dependence of $\tau$ on $T$ (Eq.\,\ref{eq:tau-hopping}).

In the work of B\"osch (\cite{Bosch78}), Fitzgerald et al.
(\cite{Fitzgerald01}) and others, a comparison between experimental and
calculated temperature dependancies was performed .  It showed that
relaxation processes should be taken into account for temperatures $T
\ga 10\,K$.  At these temperatures, relaxation dominates over resonant
tunneling.  Hopping relaxation becomes significant at $T$ about 25\,K
and dominates at higher temperature.  This tendency has a general
character, and shows a similar behavior for substances as different as
fluoride mixed crystals and silica glasses.

%



\section{Model of the dust FIR/mm emission}
\label{s:astro-model}

The theoretical considerations given in
Sect.\,\ref{Sec:ModelPhysics} provide the formalism for calculations
of dust emissivity in the far-infrared and millimeter-wave region
and can be gathered into a new model for explaining the astronomical
emission in this spectral range. So far, modeling the astronomical
dust emission spectra in this spectral range has been performed on
purely phenomenological grounds with little connection to solid
states physics.  Unlike previous models, the new approach proposed
here does not assume that dust is composed of crystalline material,
but instead uses for the first time theoretical results applicable
to disordered materials which are likely to compose astronomical
dust particles.


\subsection{Model description}
\label{s:astro-region}

The mass absorption coefficient of
dust entering Eq.\,\ref{eq:I} can be expressed as

\begin{equation}
\kappa_{\rm dust} = \kappa_\IR+\kappa_\DCD+\kappa_\TLS.
\end{equation}

The first term $\kappa_\IR$ includes the opacity in infrared bands and
other relatively short-wave absorption processes.  References to dust
models describing $\kappa_\IR$ are given in Sect.\,\ref{s:intro}.  IR
bands in the dust spectrum are linked to the optical modes of
lattice oscillations, like bending vibration bands of interatomic bonds
such as \mbox{O-Si-O} or \mbox{H-O-H}.  As was shown in the previous
sections, the long wavelengths wing of infrared bands is relatively
weak in the FIR/mm.  Our model does not provide a new description of
these processes and is complementary to IR models.  While
$\kappa_\IR$ depends mainly on the chemical composition of dust, the
following terms $\kappa_\DCD$ and $\kappa_\TLS$ are linked to the
disordered structure of the grain material and are largely
independent of the dust chemical composition.

As described in Sect.\,\ref{Sec:ModelPhysics}, $\kappa_\DCD$ and
$\kappa_\TLS$ are linked to the existence of disorder in the structure
of the grain material.  Given the fast decrease of $\kappa_\IR$ with
wavelength, these effects are likely to dominate absorption in the
FIR/mm. Disordered media absorption takes place in all substances excluding
perfect crystals.  Amorphous dust as well as partially ordered or dirty
crystalline dust can therefore be described by the model.

The first term, $\kappa_\DCD$ corresponds to disordered charge
absorption described in Sect.\,\ref{sec:mex}.  It is temperature
independent. The second term,
$\kappa_\TLS$ is connected with
transition of charges in a lattice between vacant potential minima due
to tunneling or thermal activation (Sect.\,\ref{Sec:TLS}).  This term
displays a specific spectrum of absorption over a wide range of
frequencies $\omega/2\pi$, from about 1\,Hz to about $1$\,THz.  It
includes three components designated in the model as resonant
($\kappa_{\rm res}$), tunneling relaxation ($\kappaphon$), and
hopping relaxation ($\kappa_{\rm hop}$) respectively :

\begin{equation}
\kappa_\TLS(\omega,T) =
\kappa_{\rm res}(\omega,T)+\kappaphon(\omega,T) +\kappa_{\rm hop}(\omega,T).
\label{eq:kTLS}
\end{equation}

Spectral index of $\kappa_\TLS$ significantly deviates from
$\beta=2$ and depends on temperature. Resonant opacity $\kappa_{\rm res}$
decreases with $T$, relaxation opacity $\kappaphon$ and $\kappa_{\rm hop}$
rises with $T$.

The various opacities of the model can be summarized as,

\begin{eqnarray}
\kappa_\DCD (\omega) &=& K_\DCD \omega^2
   \left[1-\left(1+\omega^2/\omega_c^2\right)^{-2}\right]\label{eq:k-res},\\
\kappa_{\rm res} (\omega,T) &=& K_\TLS \omega\
   g_{\rm res}\left(\frac{\omega}{\omega_{\rm m}}\right)\tanh \left(\frac{\hbar\omega}{2 kT} \right),\\
\kappaphon (\omega,T) &=&\frac{1}{4 \pi^2}K_\TLS\ \omega\ \Fphon
(\aphon \omega T^{-3}),\\
\kappa_{\rm hop} (\omega,T) &=&
\frac{2}{\pi} K_\TLS \omega  (\cD+\ln T)
\int^{\infty}_{V_{\min}} \frac{\omega\tau(V,T) P(V)\ud
V}{1+\omega^2\tau^2(V,T)}\label{eq:k-hop},
\end{eqnarray}

where $\tau(V,T)=\tau_0\ \exp(V/\kB T)$.

The amplitude factors for the DCD and TLS terms can be expressed with
respect to material properties as follows:

\begin{eqnarray}
K_\DCD &=& \overline{q^2_{\rm m}}\ (3\upsilon_t^3 c\rho)^{-1}, \label{eq:KDCD}\\
K_\TLS &=& \frac{4}{3} \pi^2\overline{P}\mu_{\rm b}^2\ (c\rho)^{-1}, \label{eq:KTLS}\\
\aphon &=& \pi\gamma_{\rm e}^{-2}\upsilon_t^5 \rho \hbar^4 (2\kB)^{-3}.
\end{eqnarray}

The function $g_{\rm res}$ is the normalized ODOS
spectrum given by Eq.\,(\ref{eq:ODOS}). The function $F_{\rm
rel}$ is defined by Eq.\,(\ref{eq:varxy}) and can be calculated
interpolating Eq.\,(\ref{eq:interpol}) using precalculated values in
Table\,\ref{tab:f2} of the Appendix.

The Gaussian probability distribution function of the TLS potential
barrier height $P(V)$ is defined by
Eq.\,(\ref{eq:Gauss-trunk}) with parameters $V_{\rm m}$, $V_0$ and $V_{\min}$.

$K_\DCD$ and $K_\TLS$ define the amplitude scale of $\kappa_\DCD$ and
$\kappa_\TLS$.  Parameters $\omega_c$, $\omega_{\rm m}$, $\aphon$,
$\tau_o$ define the scale of $\kappa_\DCD,\ \kappa_{\rm res},\
\kappaphon,\ \kappa_{\rm hop}$, respectively, along the $\omega$
axis.

The numerical values and units of the various parameters entering the model are
given in Tab.\,\ref{tab:params}. They set the frequency and temperature
dependence of the model.

It is important to note that the values proposed here only give an
order of magnitude for the parameters, and are derived only from laboratory
measurements (Sect.\,\ref{Labevidences} and \ref{Sec:ModelPhysics}) or
theoretical expectations. They provide reference values only. They
are not tuned to reproduce observed astronomical spectra. Derivation
of the parameter values to be used for astronomical purposes will be derived
in a forthcoming paper.

\begin{table}[ht]
\caption[]{\label{tab:params} Parameter values.}
\begin{flushleft}
\begin{tabular}{lll}
\hline
\hline
parameter & value & unit\\
\hline 
$K_\DCD$$^\dag$   & $1.2\E{-24}$ & ${\rm cm^2g^{-1}s}^2$\\
$K_\TLS$$^\dag$   & $1.5\E{-13}$ & ${\rm cm^2g^{-1}s}$\\
$\omega_c$$^\ddag$ & $3.10^{12}$    & ${\rm s}^{-1}$\\
$\omega_{\rm m}$$^\flat$ & $2.7\E{12}$ & ${\rm s}^{-1}$ \\
$\aphon$$^\dag$ & $5.3\E{-10}$ & ${\rm K^3s}$\\
$\tau_o$ $^{\star\star}$ & $10^{-13}$ & ${\rm s}$\\
$V_{\rm m}$ $^\diamond$ & $7.6\E{-14}$ & ${\rm erg}$ \\
$V_0$ $^\diamond$ & $5.7\E{-14}$ & ${\rm erg}$ \\
$V_{min}$ $^\diamond$ & $6.9\E{-15}$ & ${\rm erg}$ \\
$\cD$ $^\star$& $5.8$ & $$ \\
\hline
\hline
\end{tabular}
\end{flushleft}
$^\dag$ values calculated using Eq.\,(\ref{eq:KDCD}-100) and the
following approximate mean values : \mbox{$\upsilon_t= 3\E5$}\,cm/s,
$\rho$ = 3 g/cm$^3$, $\overline{P}\mu_{\rm b}^2=10^{-3}$, $\gamma_e$
= 1\,eV, $\overline{q^2_{\rm m}}=e^2/m_o$, where $e$ is the
electron charge and $m_o$ is the mass of the Oxygen atom.\\
$^\ddag$ using $l_c=\upsilon_t/\omega_c$=1 nm (see
Tab.\,\ref{tab:par})\\
$^\flat$ corresponds to $2\pi c/\omega_{\rm m}$=0.7\,mm (see
Sect.\,\ref{s:res})\\
$^\diamond$ from Table\,\ref{tab:par}\\
$^{\star\star}$ $\tau_o$ from Sect.\,\ref{s:relax}\\
$^\star$ evaluated from Sect.\,\ref{s:hopping}.
\end{table}



\subsection{Model applicability}

\subsubsection{Wavelength range of validity}

The model presented here applies to FIR to millimeter emission from
amorphous materials. In the astronomical context, it should be used
to describe the emission of large dust grains which usually dominate
the observed spectra in the FIR/mm. In principle, it should also
apply to the emission from amorphous very small 3-D particles which
emission usually dominates the mid-IR region of astronomical
spectra.


The DCD component is valid only in the frequency range dominated by
acoustic oscillations.  Towards high frequencies the corresponding
limit is set where the sound wavelength is of the order of the
interatomic distances $d$. For silicates this corresponds to an
electromagnetic wavelength $\lambda_d \simeq 10\mic$.  At shorter
wavelengths IR models should be used. Towards the long wavelengths,
the validity is limited to $\omega \leq \omega_a$  with

\begin{equation}
\omega_a \approx \pi\upsilon_t/a,
\end{equation}

where $\upsilon_t$ is the transverse sound velocity and $a$ is the
grain size. The DCD model should therefore be valid for all wavelengths
below $\lambda_{a}=3\,mm$ for the range of sizes usually assumed for
the BG component (e.g. $15-110\,nm$ in D\'esert et al. \cite{Desert90}), using
$\upsilon_t=3\,10^{5}\,cm/s$. Application of this part of the model
to small transiently heated dust particles in the long wavelength
range should therefore be handled with care.




The TLS part of the model is in principle not limited to a given
spectral range.

\subsubsection{Applicability in dust models}

In general, modeling FIR/mm dust emission from an interstellar source
 be seen as a three-step process. The first step is to select the materials
which constitute the grains, and specify the optical
properties (complex dielectric constants or refractive index, or
opacities) of the selected materials. The second step is to chose, for
each dust component, a distribution of size, shape, composite
structure, orientation with respect to the direction of the
incoming radiation field, in order to calculate
extinction, absorption and diffusion using Mie or some multipole
approximation method. The third step is to specify the
dust density and radiation field distributions, and to include
radiative transfer to calculate the dust emission on a given line of
sight.
The model presented here concerns the first step of this dust
emission modeling only.

The equations proposed in Sect.\,\ref{s:astro-model} do not include
the local field correction factor given in
Eq.\,(\ref{eq:Aabs}).
They are valid for spherical or quasi-spherical particles.  Any
possible effect due to anisotropic shapes, such as a polarized
emission by grains for example, is beyond the scope of this paper,
and refers to the second step of modeling.

The model equations do not depend on macroscopic values of the
dielectric constant and average density. All the parameters involved
in the model have a local, microscopic meaning. Therefore the
derived emissivities are defined by the total mass of dust, and do
not depend on the porosity of the grains.  The model should
therefore be equally valid for fluffy and for bulk grains.

It is important to note that unlike standard modeling, our model
shows that it is not sufficient for the first modeling step to
select from databank some complex dielectric constants
$\epsilon(\omega,T)$ obtained for bulk mineral samples of chemical
composition corresponding to interstellar element abundancies.
First, the particle size should be taken into account. Indeed, if
the opacity (mass emissivity) of dust (Eq.\,\ref{eq:k-eps}) does not
depend on particle size for a given dieletric constant
$\epsilon(\omega)$, the dieletric constant$\epsilon(\omega)$ itself
is size dependent, as shown in Sect.\,\ref{s:astro-region}.
Second, both DCD and TLS processes are more sensitive to the degree
and the type of disorder in a given material, than to
the exact chemical composition of the grain.

\subsection{ Temperature and frequency dust opacity behavior}\label{s:astro-behavior}

In the general case, the model can produce a wide variety of
spectral shapes, ranging from a "classical" spectrum $\beta$=2 when
only DCD is taken into account ($K_{TLS}=0$) and the correlation
length caracterizing DCD is assumed infinite ($\omega_{c}$=0) to
spectra with much flatter behaviors when TLS effects are included,
with mean spectral index values $\beta$ over the whole FIR/mm range
close to one at high dust temperatures. Some extreme cases even
include the possibility of having $\beta<0$ or $\beta>2$ over some limited
regions of the spectrum, due to the influence of the resonant TLS
absorption and to the DCD cut-off respectively.

However, the model generates a "typical emission profile" when a
correlation length of nanometric scale is assumed, and when the TLS
effects are expected and taken into account. This emission profile
presents some interesting characteristic deviations compared to the
commonly adopted profile (a modified black-body type emission
characterized by a constant spectral index $\beta$=2). To illustrate
this fact, we show on Fig.\,\ref{f:absor} the dust opacity spectrum
including both the DCD and TLS components ($\kappa_{\rm dust} =
+\kappa_\DCD+\kappa_\TLS$), calculated using
Eq.\,(\ref{eq:k-res}-\ref{eq:k-hop}) and adopting the parameter
values given in Tab.\,\ref{tab:params}. First, the slope of the dust
opacity versus wavelength, and consequently the spectral index, are no
longer constant over the whole FIR/mm spectral range.
As the temperature increases, it begins to deviate in the longest
wavelength range, starting from a local spectral index values higher
than $\beta$=2 around T=10K, towards lower values down to around
$\beta$=1. Second, as the temperature increases more, this spectral
behavior propagates towards the short wavelengths of the FIR/mm
range. It is remarkable that only with model parameter values
referenced in the solid state physics literature, the
temperature-dependent spectral behavior of the modeled dust opacity
is in qualitative agreement both with laboratory data in that
temperature range (Sect.\,\ref{sec:Sch} and \ref{Labevidences}).

\begin{figure}
\includegraphics[width=8cm]{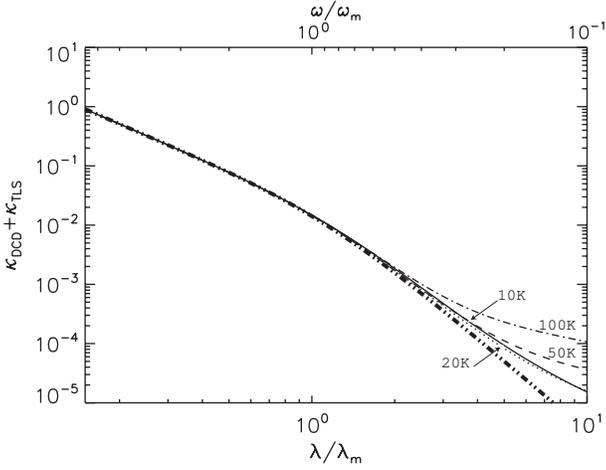}
\caption{\label{f:absor}
Total dust opacity spectra $\kappa_\DCD+\kappa_\TLS$.  The spectra are
given in arbitrary units, normalized at $\lambda=100\mic$ and $T=10\,K$.  The
dash-dot-dot line shows the temperature independant DCD dust opacity
alone computed for $\lcor=1\,nm$.  Line styles for the temperature
dependant total effect are the same as in Fig.\,\ref{fig:kmI}.
}
\end{figure}

%

For each individual components of the model, the temperature and
frequency behaviors have been described in
Sect.\,\ref{Sec:ModelPhysics}. The DCD opacity is temperature
independent and shows a quadratic frequency behavior
($\kappa_\DCD\sim\omega^2$) for high frequencies and drops
dramatically ($\omega<\omega_c$) at lower frequencies. The
transition between the two regimes is set by the value of the
correlation frequency ($\omega_{c}$), or equivalently the
correlation length of the charge distribution in the grain
material.  We point out that, although the transition between
these two regimes is very smooth, this distinctive behavior opens
the possibility to measure the correlation length from astronomical
emission spectra. Note that the DCD effect is the only one with such
a rapid fall-off of the absorption with wavelength, which seems to
be measured toward some astronomical sources which exhibit large
$\beta$ values, including values higher than 2.

Resonant tunneling opacity takes place around $\omega<\omega_{\rm
m}$ and dominates at low temperatures since $\kappa_{\rm
res}\sim\omega^2/T$ (Fig.~\ref{fig:kmI}).  As was shown in
Sect.\,\ref{sec:mex}, the wavelength $\lambda_{m}=2\pi c/\omega_{\rm m}$
is in the spectral range from $500\mic$ to $1\,mm$ for
silicates.

The tunneling relaxation spectrum is closely linked to the resonant
tunneling. Their added effect is shown on Fig.~\ref{fig:tun}. The
amplitude of the resonant feature decreases with rising temperature,
while the amplitude of the relaxation wings increases.

\begin{figure}
\includegraphics[width=8cm]{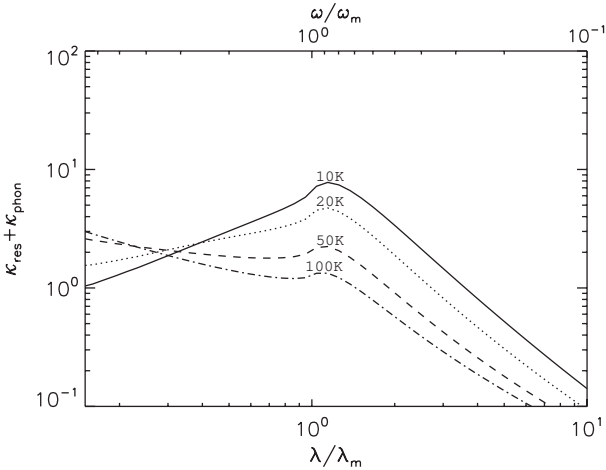}
\caption{
\label{fig:tun}
Temperature dependence of the resonant tunneling and phonon induced
tunneling relaxation opacity (See Fig.~\ref{fig:kmI} for comparison).
The spectra are given in arbitrary units, normalized at
$\lambda=100\mic$ and $T=10\,K$.  Line styles are the same as in
Fig.\,\ref{fig:kmI}.
}
\end{figure}

The hopping relaxation behaves as a distinct spectral feature in the
case of narrow barrier height distributions (Sect.\,\ref{s:hopping},
Fig.~\ref{fig:hop1Hz}).  The amplitude of the effect rises with
frequency as $\kappa_{\rm hop}\sim\omega^2$, until it reaches a
maximum at $\omega=\tau_0^{-1}\exp(-V_0/\kB T)$.  At low temperature
the location of the maximum is shifted to MHz frequencies.  It
reaches the FIR/mm region at $T \simeq 100\,K$, a value which
depends upon $V_0$.  The hopping relaxation in the case of wide
barrier height distribution, creates a very wide spectral
distribution (Fig.~\ref{fig:hop1Hz}).  The behavior is then
$\kappa_{\rm hop}\sim T\omega^\beta$ with a frequency-dependent
slope value, usually $\beta<1$ (Eq.\,\ref{eq:hop-mm},
Fig.~\ref{fig:hop-mm}).

At a given temperature and wavelength, the relative efficiency of
the TLS effects compared to the underlying DCD effect is governed by
the density and efficiency of the optically active TLS in the
material, and so by the value of the parameter $\overline{P}\mu_{\rm
b}$. This value should be related to the fluffiness of the material
and to the types and densities of defects in its structure.

Generally, the resonant part of the TLS effects is expected
to dominate over the other relaxation processes at temperature
below 10K, while the TLS hoping mechanism should fully dominate
above 30K.

\subsection{Model relevance}\label{Sec:Discussion}

Recent astronomical observations of the interstellar dust emission
in the FIR/mm wavelength range show some unexpected features which
can only be related with difficulty to previous knowledge about dust
optical properties in that range (Sect.\,\ref{s:astro}. All data
have to be explained with a blackbody emission modified by a
frequency power law, whose exponent is constant over that FIR/mm
range, with a temperature independent asymptotic value equal to two.
Indeed, very little is known about the optical properties of the
dust amorphous state in that range. Often standard synthetic dust
permitivities or opacities are used, which fit perfectly the near
and mid-infrared observed features on the basis of a sum of
Lorentz-type resonant profiles. But in the absence of reliable data and
knowledge, the FIR/mm behavior is often only modeled by the
vanishing long-wavelength wings of those infrared resonances. In
addition some laboratory data on various silicates reveal additional
absorption features, strongly temperature dependent in the FIR/mm
range.

It is clear that only specific processes with characteristic
frequencies located in the FIR/mm region are needed to explain non
purely asymptotic features in this spectral domain. Two phenomena
satisfying these requirements are known in solid state physics.
First, acoustic oscillations with acoustic wavelengths less than grain
size have frequencies in the range $3\E{10}-3\E{12}$\,Hz
(Sect.\,\ref{s:astro-region}). These acoustic oscillations are
"optically" active (interact with electromagnetic field) in
disordered media (Sect.\,\ref{sec:mex}). Second, tunneling states in
the disordered solids have frequencies $\Delta_0/h$ falling in the
frequency range $5\E6-5\E{11}$\,Hz (Sect.\,\ref{s:hopping}).

These two phenomena are linked to disordered material composing
amorphous solids and dirty crystalline compounds.  This is also the
case of interstellar grains. They dominate the FIR/mm emission of
the highly amorphous interstellar dust (Sect.\,\ref{LorentzModel}).

The transverse sound oscillations and tunneling between the
potential minima in a disorder lattice includes rotation of groups
of atoms, limited and influenced by interactions with their
neighboring atoms. In the classification of motion into electronic,
vibrational and rotational states used in fundamental quantum
mechanics, this model refers to rotational motions, when the former
dust emissivity models used only asymptotic behavior of electronic
and vibrational processes. In that sense, the model proposed here
does not replace the former ones, but complements them by taking
into account the physical phenomena efficient in the low frequency
range.

Both processes are sensitive to the degree and the type of disorder,
rather than to the exact chemical composition of the material.  As a
consequence, the FIR/mm emission profile is governed by a few
key-parameters which characterize the disorder.  There is no
contradiction between this and laboratory results of FIR/mm dust
absorption.  It is important to investigate the variations of the
model parameters in laboratory measurements of various duly
characterized samples.  For this purpose, the model gives new
avenues for such simulations, and such disorder characterization. It
is likely that the FIR/mm properties will be more influenced by the
type and degree of disorder than by the chemical composition.

It is important to note that the  parameter values adopted in this
paper were chosen from purely theoretical grounds or from laboratory
results obtained on materials that may not be exactly representative
of ISM dust particles.  We make no attempt in this paper to model
the actual astronomical observations.  This will be performed in a
companion paper.
It is important that only the later parameters be used for
astronomical purpose. As the type and degree of disorder in
interstellar dust should be representative of the amorphous dust
formation itself, we hope that this model will open the way to bring
new insights about the interstellar medium dust.

\section{Conclusion}
\label{Sec:Conclusion}

The continuum FIR/mm emission is generally attributed to an
interstellar dust component of amorphous grains, silicate-based,
large enough ($\simeq 15-100\,nm$) to radiate at thermal equilibrium. So far
the conception of FIR/mm dust emissivity was built on the
semi-classical Lorentz or Drude models. They lead in the long
wavelength limit to a dust emissivity obeying to a frequency power
law, whose exponent (the spectral index) equals 2 and is usually
considered as temperature independent. However a growing volume of
FIR/mm observational data induces some doubts about the
applicability of such models in that wavelength range.

In this context we present a new model for FIR/mm dust emission,
based on physical properties of the disordered matter.  We consider
the interaction between the electromagnetic field and the acoustic
oscillations in a Disordered Charge Distribution and a distribution
of low energy Two Level Tunneling states.  Both mechanisms apply to
the amorphous materials with a large degree of universality,
independant of the exact chemical nature of the dust.

The proposed model predicts a disordered induced FIR/mm dust emission
which dominates over the weak long wavelength wings of the infrared
resonances.  This emission is strongly temperature dependent, and leads
to a millimeter emission enhancement relative to predications from
classical models.  The FIR/mm emission spectra can
not longer be characterized by a constant spectral index over that
range.  The detailed shape of the emission spectra is governed by a
few key parameters which characterize disorder.  The exact values
of these parameters for interstellar dust should be determined from
observations.

Our model is complementary to the classical models which are reliable
in the infrared, and adds emission processes due to
disorder induced features.


\begin{acknowledgements}

Part of this work was supported by Egide and INTAS grants 01-686 and
05-02-19650, and the french national program PCMI. The PhD thesis
work of D. Paradis was supported by the "Fond Social Europeen"
European grant and the Noveltis company.

\end{acknowledgements}

\appendix
\section{TLS optical density of states}
\label{ap:ODOS}

The ODOS spectrum for continuous distribution of DOS
is calculated by integration of Eq.\,(\ref{eq:G-p}) over the energy
plitting $\Delta_0$. We use the following polynomial form of the
distribution function $\pG(x)$, with $x=\Delta_0/\hbar\omega_{\rm m}$

\begin{equation}
    \label{eq:pol-DOS}
    \pG(x) = \left\{
    \begin{array}{ll}
       \frac{7}{8}(1+3x^2)(1-x^2)  &\mathrm{for} \ x<1,\\ 
       \hspace{15mm} 0 &\mathrm{for} \ x>1, 
    \end{array} \right.
\end{equation}

Equation\,(\ref{eq:pol-DOS}) is a least square polynomial fit to
Eq.\,(\ref{eq:Pconst}), removing the discontinuity at x=1.  A similar
result is obtained using the first Fourier harmonics of
Eq.\,(\ref{eq:Pconst}).

The calculated ODOS spectrum is given by
$G(\omega)=\overline{P}\mu_{\rm b}^2 g(x)$ with $x=\omega/\omega_{\rm m}$
and

\begin{equation}\label{eq:g12}
    g(x) =  \left\{
    \begin{array}{ll}
        1+x^2 g_1(x) & \mathrm{for} \ x<1,\\
        1+x^2 g_1(x)-g_2(x)\sqrt{1-x^{-2}} & \mathrm{for} \ x>1,
    \end{array}\right.
\end{equation}

with

\begin{eqnarray}
&&   g_1(x) = 4(5-6x^2)/15,\label{eq:g1}
\\&& g_2(x) = 8(1-x^2)(2+3x^2)/15.\label{eq:g2}.
\end{eqnarray}

In the low frequency approximation ($\omega\ll\omega_{\rm m}$)
we have $\pG \approx 1$ and the ODOS function $G(\omega)$ does not
depend on $\omega$, in agreement with previous results of the TLS theory.

\section{Evaluation of simplified equation
for tunneling relaxation calculation.}
\label{ap:tun}
A use of simplified single integral expression of B\"osch (Eq.\,\ref{eq:B78})
instead of full double integral equation used here (Eq.\,\ref{eq:FC01})
is based on assumption that income from asymmetric states is negligible.
This approach gives correct result for resonant absorption
(Eq.\,\ref{eq:G-p}) where
integrand term $(1-\Delta_0/\hbar\omega)^{-1/2}$ amplify weight of symmetric
states $\Delta_0\approx\hbar\omega$. Another situation takes place for
relaxation absorption (Eq.\,\ref{eq:FC01}) where integrand term
$(1-\tau_1/\tau)^{+1/2}$ emphasize weight of asymmetric states $\tau\gg\tau_1$.
Direct calculations give discrepancy of results more than hundred times.
A comparison could be made without detailed calculation be consideraton
of noncoinsiding terms in  integrand expressions in integrals over d$E$.
In case of (Eq.\,\ref{eq:B78}) it is
$f_1(\omega,\tau) = \omega^2\tau_1/(1+\omega^2\tau_1^2).$\\
In case of (Eq.\,\ref{eq:FC01})
\begin{equation}\label{eq:intJ}
f_2(\omega,\tau)=\int^{\infty}_{\tau_1}\sqrt{1-\frac{\tau_1}{\tau}}
\frac{\omega^2\ud\tau}{(1+\omega^2\tau^2}.
\end{equation}
It is not difficult to define the asymptotics of both functions
$$f_1(\omega,\tau_1) \approx \left\{
\begin{array}{ll} 
        \omega^2\tau_1 &  \mathrm{for} \ \omega\tau_1 \ll 1\\
        \tau_1^{-1} & \mathrm{for} \ \omega\tau_1 \gg 1
\end{array} \right.$$
$$  f_2(\omega,\tau_1) \approx \left\{
\begin{array}{ll} 
      \omega\pi/2 &  \mathrm{for} \ \omega\tau_1 \ll 1\\ 
      (2/3)\tau_1^{-1} & \mathrm{for} \ \omega\tau_1 \gg 1
\end{array} \right.$$
The asymptotic behaviors of $f_1$ and $f_2$ are different
and the simplified equation is not accurate also for wide region
of values of parameters.

\section{Calculation of integral for tunneling relaxation spectrum.}

The function $\Fphon (\omega,T)$ (Eq.\,\ref{eq:f2wT}) can be simplified
into a function of a single parameter $\Fphon (p)$ with $p=
a\omega/(2\kB T)^3$ through a change of variables to the dimensionless
variables $x=\arctan(\omega\tau)$ and $y=\tanh(E/2\kB
T)$. Equation\,\ref{eq:f2wT} then writes

\begin{equation}\label{eq:varxy}
\Fphon (p) = \int^1_0\int^{\pi/2}_{x_1} \sqrt{ 1-\frac{\tan x_1}{\tan x} }\ \ud x \ud y,
\end{equation}

where

\begin{equation}
x_1 = \arctan( p\ y\textrm{ arcth}^{-3} y),
\end{equation}


The function $\Fphon(p)$ can be further approximated using a fit of
the form,

\begin{equation}\label{eq:interpol}
\Fphon (p)=\Fphon (p_i)\left(\frac{p}{p_i}\right)^{\beta_{2\, i}},\ p_i<p<p_{i+1}.
\end{equation}

Precalculated values of the function $\Fphon$ and the coefficients
$\beta_2$ entering the interpolated form are given in
Tab.\,\ref{tab:f2}.

\begin{table}[ht]
\caption[]{\label{tab:f2} Values of $\Fphon (p_i)$ and $\beta_2(p_i)$
entering the interpolated form of $\Fphon(p)$ in Eq.\,\ref{eq:interpol}.}
\begin{flushleft}
\begin{tabular}{lll}
\hline
\hline
$p_i$ & $\Fphon(p_i)$ & $\beta_2(p_i)$\\
\hline 
0.001 & 1.4696 & -0.0321\\
0.002 & 1.4313 & -0.0446\\
0.005 & 1.3604 & -0.0678\\
0.01 & 1.2875 & -0.0924\\
0.02 & 1.1948 & -0.1248\\
0.05 & 1.0393 & -0.1826\\
0.1 & 0.8984 & -0.2397\\
0.2 & 0.7433 & -0.3090\\
0.5 & 0.5334 & -0.4185\\
1 & 0.3866 & -0.5114\\
2 & 0.2623 & -0.6084\\
5 & 0.1417 & -0.7343\\
10 & 0.0826 & -0.8208\\
20 & 0.0455 & -0.8950\\
50 & 0.0194 & -0.9659\\
100 & 0.0098 & -0.9897\\
200 & 0.0049 & -0.9969\\
500 & 0.0020 & -1.0000\\
1000 & 0.0010 & \ldots\\
\hline
\hline
\end{tabular}
\end{flushleft}
\end{table}


\section{Calculations of integral for hopping relaxation}\label{a:hopping}

Following the TLS formalism, the hopping relaxation spectrum can be
computed as the integral
of Eq.\,(\ref{eq:Debye-Phill}) over a distribution of TLS and barrier
heights,
taking into account that  in contrast to tunneling relaxation,
the time constant $\tau$ depends on barrier height $V$ according to Eq.\,
(\ref{eq:hopping}) and does not depend directly on TLS parameters $\Delta_0$ and $\Delta$:

\begin{eqnarray}
\label{eq:triple}
&&\chi_0''(\omega)=\frac{\mu_{\rm b}^2}{3\kB T}
\int_0^\infty\int_{-\infty}^{+\infty}\int_0^\infty\frac{\overline{P}}{\Delta_0} P(V)\times
\nonumber\\&\times&
\frac{\Delta^2}{E^2}\sech^2\Big(\frac{E}{2\kB T}\Big)
\frac{\omega\tau}{1+(\omega\tau)^2}\ \ud\Delta_0\ \ud\Delta\ \ud V.
\end{eqnarray}

For calculating this expression, the convention of Hunklinger and Schickfus
(\cite{Hunklinger81}) is used, which allows to reduce the triple integral to
a single
integration over distribution $P(V)\ud V$, where $P(V)$ is taken to
be Gaussian.
This convention suggests the independence of the distribution $P(\Delta_0,\Delta)$ and
$P(V)$,
which permits to separate the integrals and transforms Eq.\,(\ref{eq:triple})
into
Eq.\,(\ref{eq:hopping}),
in a form similar to that of Fitzgerald et al. (\cite{Fitzgerald01}). In this case $P(V)$ is
an independent Gauss probability distribution following the standard normalization
$$\int_0^\infty P(V)\ \ud V=1$$

which defines the  normalization coefficient $C_V$ in Eq.\,(\ref{eq:Gauss-P}) as

\begin{equation}\label{eq:norm}
C_V=\frac{1}{V_0\sqrt{\pi}}
\Big[\frac{1}{2}{\rm Erf}\Big(\frac{V_{\rm m}-V_{\min}}{V_0}\Big)+\frac{1}{2}\Big]^{-1},
\end{equation}

where the term in brackets is usually about unity.

The coefficient $B_{\rm hop}(T)$ in Eq.\,(\ref{eq:hopping}) is equal to

\begin{eqnarray}\label{eq:hopping-2}
&&B_{\rm hop}(T)=\frac{4\pi}{c\sqrt{\epsilon'}}\frac{(\epsilon'+2)^2}{9}
\frac{2}{3}\overline{P}\mu_{\rm b}^2
\nonumber\\&&\times\nonumber
\int_{\Delta_0^{\min}}^\infty\int_{\Delta_0^{\min}}^E
\sech^2\Big(\frac{E}{2\kB T}\Big) \sqrt{1-\frac{\Delta_0^2}{E^2}}
\frac{\ud\Delta_0}{\Delta_0} \frac{\ud E}{\kB T}=
\\&&=\frac{4\pi}{c\sqrt{\epsilon'}}\frac{(\epsilon'+2)^2}{9}\frac{2}{3}\overline{P}\mu_{\rm b}^2
(\ln\frac{\kB T}{\Delta_0^{\min}}+C_1),
\end{eqnarray}

where the constant $C_1$ is given by

\begin{equation}
C_1=\ln 4-1+\int_0^1\ln{\rm arcth}\ x\ \ud x\ =\ -0.441. \label{eq:hopping-C}
\end{equation}

The integral (\ref{eq:hopping-2}) over d$\Delta_0$ has been
determined precisely. Integrating over d$E$ was performed assuming
that terms of order $\Delta_0^{\min}/\kB T$ and higher orders could
be neglected, taking into account that $\Delta_0^{\min}/\kB T\ll 1$.



\begin{thebibliography}{}


\bibitem[1996]{Agladze96}
Agladze N.I. et al. 
1996, \apj, 462, 1026

\bibitem[1998]{Agladze98} Agladze N.I., Sievers A.J. 1998,
\prl, 80, 4209

\bibitem[1994]{Agladze94} Agladze N.I., Sievers A.J., Jones
S.A., Burlitch J.M., Beckwith S.V.W.  1994, \nat, 372, 243

\bibitem[1985]{Allamandola85}Allamandola L.J., Tielens A., Barker J.R.
1985, \apj, 290, L25

\bibitem[1972]{Anderson72}Anderson P.W., Halperin B.I., Varma C.M. 1972,
Phil. Mag., 25, 1

\bibitem[1993]{Andre93}Andre P. et al. 1993, 
\apj, 406, 122A

\bibitem[1974]{Andriesse74} Andriesse C.D. 1974, \aap 37, 257

\bibitem[2006]{Barreiro04}Barreiro, R. B. et al., 
2006, \mnras, 
in press 

\bibitem[2003]{Bennett03}Bennett, C. L. et al. 2003, 
\apjs, 
148, 97

\bibitem[1999]{Bernard99}
Bernard J.-P. et al. 
1999, \aap, 347, 640

\bibitem[1978]{Bosch78} B\"osch M.A.
1978, Phys.~Rev., 40, 879

\bibitem[1967]{Bosomworth67} Bosomworth D.R.,
1967, Phys. Rev., 157, 709

\bibitem[2002]{Boudet02}Boudet N. et al. 
2002, in Chemistry as a Diagnostic of Star Formation,
ed. C. Curry and M. Fich (NRC Press, Ottawa, Canada), 257

\bibitem[2005]{Boudet05}Boudet N., Mutschke H., Nayral C., Jäger C., Bernard J.-P., Henning T., Meny C.,2005,
\apj, 633, 272

\bibitem[1999]{Bouchet99}Bouchet, F. R., Gispert, R. 1999, \na, 
4, 443 

\bibitem[2004]{Brucato04}Brucato J.R., G. Strazzulla et al. 2004,
\aap, 413, 395

\bibitem[1995]{Chandler95}Chandler C.J. et al. 1995, 
\apj, 455L, 93

\bibitem[1914]{Debye14} Debye P. 1914, Ann. Physique 43, 49

\bibitem[1990]{Desert90}Desert F.-X., Boulanger F., Puget J.L.
1990, \aap, 237, 215

\bibitem[1985]{Draine85}Draine B.T., Anderson N.
1985, \apj, 292, 494

\bibitem[1984]{Draine84}Draine B.T., Lee H.M.
1984, \apj, 285, 89

\bibitem[2001]{Draine01}Draine B.T., Li A.
2001, \apj, 551, 807

\bibitem[2001]{Dupac01G}
Dupac X. et al. 
2001, \apj, 553, 604

\bibitem[2002]{Dupac02}
Dupac X. et al. 
2002, \aap, 392, 691

\bibitem[2003]{Dupac03}
Dupac X., Bernard J.-P. et al. 
2003, \aap, 404, L11

\bibitem[2003]{Dupac01B}
Dupac X., del Burgo C. et al. 
2003, \mnras,  344, 105

\bibitem[1997]{Dwek97}
Dwek E. et al. 
1997, \apj, 475, 565

\bibitem[1922]{Ewald22} Ewald P. P., 1922, Naturwiss. 10,
1057

\bibitem[1999]{Finkbeiner99} Finkbeiner D.P., Davis M., Schlegel D.J.
1999, \apj, 524, 867

\bibitem[1994]{FitzGerald94} Fitzgerald S.A., Campbell J.A.,
Sievers A.J., 1994, Phys. Rev. Lett. 73, 3105

\bibitem[2001a]{Fitzgerald01a} Fitzgerald S.A., Sievers A.J., Campbell J.A.
2001a, J. Phys. Condens. Matter, 13, 2177

\bibitem[2001b]{Fitzgerald01} Fitzgerald S.A., Campbell J.A., Sievers A.J.
2001b, J. Phys. Condens. Matter, 13, 2095

\bibitem[2003]{Galliano03} 
Galliano F., Madden S.C., Jones A.P., Wilson C.D., Bernard J.-P.
Le Peintre F., 2003, \aap, 407, 159

\bibitem[2005]{Galliano04}
Galliano, F., Madden, S. C., Jones, A. P., Wilson, C. D., \& Bernard, J.-P. 2005,
\aap, 434, 867

\bibitem[1988]{Gordon88}Gordon M.A. 1988, %
\apj, 331, 509

\bibitem[1990]{Gordon90}Gordon M.A. 1990, 
\apj, 352, 636

\bibitem[2002]{Gromov02}Gromov V. et al. 2002,
in 
Experimental Cosmology at mm-waves,
ed. M. De Petris, M. Gervasi,
Amer. Inst. of Phys. Conf. Proc. Ser., 616, 205

\bibitem[1970]{Hadni70} Hadni A., 1970, in S.S. Mitra and S.
Nudelman (eds.), Plenum Press, New York, 561

\bibitem[1997]{Henning97}Henning T., Mutschke H.
1997, \aap, 327, 743

\bibitem[1997]{Heuer97}Heuer A.
1997, \prl, 78, 4051

\bibitem[2003]{Hubbard03} Hubbard B.E. et al. 
2003, \prb, 67, 144201

\bibitem[1981]{Hunklinger81} Hunklinger S., Schickfus, M. v., 1981,
in Amorphous Solids: Low Temperature properties,
ed. W.A. Phillips (Springer-Verlag), 81

\bibitem[1972]{Jackle72}J\"ackle J.
1972,  Z. Physik A, 257, 212

\bibitem[2003]{Jager03}J\"ager, C.D. Fabian et al. 2003,
\aap, 401, 57

\bibitem[2004]{Kemper04}Kemper, F., W. J. Vriend, et al. 2004,
\apj, 609, 826

\bibitem[1980]{Koike80}Koike C., Hasegawa H., Manabe A.
1980. \apss,  67, 495

\bibitem[2003]{Kuhn01}K\"uhn R. 2003,
Europhys. Lett., 62, 313

\bibitem[1994]{Lamarre94}Lamarre J.-M., 1994,
Infrared Phys. and Techn., 35, 277

\bibitem[2003]{Lamarre03}
Lamarre J.M. et al. 
2003,  \nar,  47, 1017

\bibitem[1982]{Landau82}Landau L.D., Lifshits E.M. 1982,
Course of Theoretical Physics, vol. VIII (Moscow, Nauka)

\bibitem[1984]{Leger84}Leger A., Puget J.L.
1984, \aap, 137, L5

\bibitem[1997]{Li97}Li A., Greenberg J.M.
1997, \aap, 323, 566

\bibitem[1977]{Mathis77}Mathis J.S., Rumpl W., Nordsieck K.H.
1977, \apj, 217, 425

\bibitem[1998]{Mennella98}Mennella V. et al.
1998, \apj, 496, 1058


\bibitem[1975]{Mon75} Mon K.K., Chabal Y.J. and Sievers
A.J., 1975, \prl, 35, 1352

\bibitem[2005]{Naselsky05}Naselsky, P. D. et al.,
2005, Int. J. Mod. Phys. D, 
14, 1273

\bibitem[1994]{Oldham94}Oldham P. et al. 1994, %
\aap, 284, 5590

\bibitem[2006]{Pajot06}Pajot F., 2006, \aap, 447, 769

\bibitem[1972]{Phillips72}Phillips, W. 1972,
J. Low Temp. Phys., 11, 757

\bibitem[1987]{Phillips87} Phillips W.
1987, Rep. Prog. Phys., 50, 1657

\bibitem[1968]{Rast68} Rast H.E., Caspers H.H., Miller S.A.
1968, Phys. Rev. 171, 1051

\bibitem[1995]{Reach95} Reach W.T., Dwek E., Fixsen D.J., Hewagama T.
et al., 1995, \apj, 451, 188

\bibitem[1998]{Ristorcelli98}
Ristorcelli I. et al. 
1998, \apj, 496, 267

\bibitem[1912]{Rubens12} Rubens H. and Hertz G., 1912,
Berlin. Ber. 14, 256

\bibitem[1964]{Schlomann64} Schl\"omann E.
1964, \pra, 135, 413

\bibitem[1982]{Schwartz82}Schwartz P. 1982, %
\apj, 252, 589 

\bibitem[2002]{Serra02}Serra G. et al. 
2002, Advances in Space Research 30, 1297

\bibitem[1998]{Sievers98} Sievers A.J., Tu J., Agladze N.,
FitzGerald S.A., Campbell J.A., 1998, Physica B, 244, 159

\bibitem[2003]{Stepnik03P}
Stepnik B. et al. 
2003 clim.conf., 187
Cross-Calibration of PRONAOS \& ISO, 
in The Calibration Legacy of the ISO Mission (ESA SP-481)

\bibitem[2003a]{Stepnik03A}
Stepnik B. et al. 
2003, \aap, 398, 551

\bibitem[1977]{Strom77}Strom U., Taylor P.C.
1977, Phys. Rev., 16, 5512

\bibitem[1975]{Schickfus75} von~Schickfus M., Hunklinger S., Pich L.
1975, \prl, 35, 876

\bibitem[1976]{Schickfus76} von~Schickfus M., Hunklinger S.
1976, J. Phys. C, Solid State Phys., 9, L439

\bibitem[1930]{Tamm30} Tamm I. E. 1930,
Z. Physik, 60, 345

\bibitem[2000]{Tegmark00}Tegmark, M., Eisenstein, D. J., Hu, W.,  de Oliveira-Costa, A., 2000,
\apj, 
530, 133

\bibitem[1960]{Vinogradov60} Vinogradov, V. S.
1960, Fiz. Tverd. Tela, 2, 2622
(English transl. 1961, Sov. Phys. - Solid St. 2, 2338)

\bibitem[1990]{Walker90} %
Walker C. et al. 1990, \apj, 349, 515

\bibitem[1986]{Weiland86}
Weiland J.L. et al. 
1986, \apj, 306, L101

\bibitem[1989]{Woody89}Woody D. et al. 1989, %
\apj, 337, 41

\bibitem[1992]{Wright92}Wright E. et al. 1992, 
\apj, 396, L13




\end{thebibliography}
\end{document}